\documentclass[reqno,12pt]{amsart}

\usepackage{fullpage}
\usepackage{graphicx}
\usepackage{amssymb}
\usepackage{amsmath}
\usepackage{cite}
\usepackage{epstopdf}

\numberwithin{equation}{section}
\def\Eqref#1{Eq.\,\eqref{#1}}
\def\Ref#1{Ref.\,\cite{#1}}
\def\ie/{i.e.}
\def\eg/{e.g.}
\def\etc/{etc.}
\def\cf/{cf.}

\def\th{\text{th}}

\def\V{\nu}

\def\f{f}

\def\E{{\mathcal E}}

\def\I{{\mathcal I}}

\def\S{3}
\def\twoS{6}

\def\sgn{{\rm sgn}}
\def\arctanh{\,{\rm arctanh}}
\def\sech{\,{\rm sech}}
\def\cn{\,{\rm cn}}
\def\sn{\,{\rm sn}}
\def\dn{\,{\rm dn}}

\def\s{\text{s}}
\def\c{\text{c}}
\def\kdv{\text{KdV}}
\def\per{\text{per.}}

\def\figref#1{Fig.~\ref{#1}}

\def\scrpt#1{$\scriptstyle {#1}$}

\begin{document}
\allowdisplaybreaks[3]

\title{Long wavelength solitary waves in Hertzian chains\\ and their properties in different nonlinearity regimes}

\author{
Stephen C. Anco$^1$
\lowercase{\scshape{and}}
Michelle Przedborski$^{2,3}$ 
\\\\\lowercase{\scshape{
${}^1$Department of Mathematics and Statistics\\
Brock University\\
St. Catharines, ON L\scrpt2S\scrpt3A\scrpt1, Canada}}
\\\\
\lowercase{\scshape{
${}^2$Department of Physics\\
Brock University\\
St. Catharines, ON L\scrpt2S\scrpt3A\scrpt1, Canada}} 
\\\\
\lowercase{\scshape{
${}^3$Department of Applied Mathematics\\
University of Waterloo\\
Waterloo, ON N\scrpt2L \scrpt3G\scrpt1, Canada}} 
}

\begin{abstract}
Properties of solitary waves in pre-compressed Hertzian chains of particles
are studied in the long wavelength limit using a well-known continuum model. 
Several main results are obtained by parameterizing the solitary waves 
in terms of their wave speed and their asymptotic amplitude. 
First, the asymptotic amplitude is shown to be directly related to the continuum sound speed,
and the ratio of asymptotic amplitude to peak amplitude is shown to describe the degree of dynamical nonlinearity in the underlying discrete system. 
Second, an algebraic relation is derived that determines the dynamical nonlinearity ratio in terms of the ratio of the solitary wave speed to the sound speed. 
In particular, highly supersonic solitary waves correspond to highly nonlinear propagating pulses in weakly compressed systems, 
and slightly supersonic solitary waves correspond to weakly nonlinear propagating pulses in strongly compressed systems.
Third, 
explicit formulas for the physical height, width, impulse and energy of the solitary waves
are obtained in both the strongly nonlinear regime and the weakly nonlinear regime.
Asymptotic expansions are used to show that in the strongly nonlinear regime, 
solitary waves are well-approximated by Nesterenko's compacton (having the same wave speed), 
while in the  weakly nonlinear regime, 
solitary waves coincide with solitons of the Korteweg-de Vries (KdV) equation, with the same wave speed. 
All of these results are illustrated by means of exact solitary wave solutions, 
including the physically important case that models a chain of spherical particles. 
\end{abstract}

\maketitle

\begin{center}\small
email: 
mp06lj@brocku.ca, 
sanco@brocku.ca
\end{center}

\section{Introduction}\label{sec:intro}

There has been considerable interest in the study of 
one-dimensional chains of discrete macroscopic particles 
that interact by a power-law contact potential 
\cite{Nesterenko1983,Nesterenko1985,Nesterenko1994,Sinkovits1995,Nesterenko1995,Sen1996, Coste1997, Sen1998, Chatterjee1999, Hinch1999, Hong1999, Ji1999,Manciu1999a,Manciu1999b, Sen1999, Hascoet2000, Manciu2001, Nesterenko2001,Sen2001, Rosas2003, Rosas2004, Daraio2005, English2005, Nesterenko2005, Daraio2006,Job2007, Sokolow2007, Zhen2007, Porter2008, Sen2008, Herbold2009, Porter2009, Rosas2010,Starosvetsky2010,Santibanez2011,James2012, Khatri2012, Stefanov2012, Takato2012, Vitelli2012,Yasuda2017,HerboldNesterenko2012,HerboldNesterenko2013,WangWensrichOoi}. 
These potentials have the form $V=a\delta^{k+1}H(\delta)$ 
where $H$ denotes the Heaviside step function, 
$\delta$ is the dynamical overlap distance between adjacent particles, 
$a$ is a constant which depends on their material properties, 
and $k>1$ is determined by the geometry of their contact surface 
\cite{Spence1968,Johnson1985,Johnson2005,Persson2006}. 
In particular, spherical particles have $k=3/2$ \cite{Hertz1882}, while $k=2$, and $k=3$ are Hertz exponents for more complicated contact geometries
\cite{Goddard1990,Sun2011}.

One of the original motivations was the experimental discovery 
\cite{Nesterenko1983,Nesterenko1985,Nesterenko1994} 
that the dynamical strain in these discrete systems 
can exhibit solitary waves, which are propagating non-dispersive localized compressive pulses. 
The existence of such waves makes these systems useful 
for a variety of physical applications related to 
shock absorption
\cite{Nakagawa2003,Sokolow2005,Hong2005,Doney2006,Melo2006,Doney2009,Breindel2011,Przedborski2015b} 
and energy localization
\cite{Vergara2006,Daraio2006energy,Job2009,Theocharis2009,Theocharis2010,Boechler2010}.

Experimental and numerical results \cite{Nesterenko2001,Daraio2006Tunability,Sun2013}
indicate that the typical wavelength of solitary waves
compared to the size of the particles in the discrete system 
is large enough to allow the use of a nonlinear continuum model 
for studying analytical properties of the solitary waves. 
The strongest nonlinearity arises when the discrete system is 
either uncompressed, with adjacent particles being just in contact, 
or weakly compressed, with the dynamical overlap $\delta$ being 
approximately at least the size of the initial overlap. 

The nonlinear continuum model for arbitrary $k>1$ 
is given by a highly nonlinear, fourth-order wave equation
\cite{Nesterenko1994,Porter2008,Porter2009,Nesterenko2001},
which we will refer to as the long-wavelength Hertzian continuum (LWHC) wave equation. 
In this model, 
the spatial gradient of solutions of the LWHC wave equation 
locally corresponds to the total strain exhibited by pulses propagating in the discrete system.
The total strain includes a contribution from the pre-compression,
which corresponds to the asymptotic value of the spatial gradient.  
In particular, solutions whose asymptotic spatial gradient is zero
provide a model for long-wavelength pulses in a discrete system with no pre-compression,
while solutions having a non-zero asymptotic spatial gradient 
represent a model for long-wavelength pulses in a discrete system with non-zero pre-compression. 

The LWHC wave equation has a well-known explicit solution 
\cite{Nesterenko1983,Nesterenko1985,Nesterenko1994}
whose gradient is a periodic travelling wave. 
A single arch of this periodic travelling wave 
can be cut off in a sufficiently smooth fashion when $1<k<5/3$ to yield 
an exact compact nonlinear wave solution, called a compacton 
\cite{Nesterenko1983,Nesterenko1994,James2012}. 
For $k>5/3$, the cutoff is singular, 
which is sometimes not emphasized in the literature. 
Since the spatial asymptotic amplitude of the compacton is zero, 
this compacton describes a compact solitary wave 
which models a propagating strictly localized pulse 
in a discrete system with no pre-compression. 
No explicit solitary wave solutions displaying a non-zero asymptotic spatial gradient 
were known until some recent work \cite{PrzAnc}
in which we obtained explicit exact solution expressions when $k=2,3$.
(We also obtained explicit periodic travelling wave solutions, 
some of which represent propagating non-dispersive localized rarefactive pulses.)
These exact solitary waves model propagating solitary wave pulses 
in a discrete system that has an arbitrary non-zero pre-compression. 

The main purpose of the present paper is to investigate 
the qualitative and quantitative properties of solitary wave solutions of 
the LWHC wave equation in different nonlinearity regimes,
and to illustrate these properties by using the known exact solitary wave solutions
in the cases $k=2,3$,
as well as in the physically important case $k=3/2$ 
whose exact solution will be derived here. 
Several interesting results are obtained. 

First, 
the background strain given by the asymptotic spatial gradient of solitary wave solutions 
is shown to be directly related to sound speed 
as defined by the dispersion relation for the linearized LWHC wave equation. 
This relationship is used to show that all solitary waves are supersonic. 

Second, 
the ratio of the background strain to the peak strain in solitary wave solutions 
is shown to describe the degree of dynamical nonlinearity in the underlying discrete system. 
An algebraic relation is derived that determines this dynamical nonlinearity ratio in terms of the ratio of the solitary wave speed to the sound speed. 
Specifically, 
highly supersonic solitary waves correspond to highly nonlinear propagating localized pulses in weakly compressed discrete systems, 
while slightly supersonic solitary waves correspond to weakly nonlinear propagating localized pulses in strongly compressed discrete systems. 

Third, 
expressions for the height, width, impulse and energy of solitary waves are derived
and shown to depend principally on the ratio of the wave speed to the sound speed. 
The width expression, which comes from an asymptotic analysis of the tail of solitary waves, has not appeared previously in the literature. 
It yields a width approximately equal to 5 particle diameters when $k=3/2$ in the strongly nonlinear case, 
which has been experimentally verified and reported many times in the literature. 
Beyond this, the expressions for the height, width, impulse and energy yield 
scaling relations that agree with ones known to hold in discrete systems. 

Fourth, 
the properties of highly supersonic solitary waves are shown to be close to the same properties of the compacton. 
In particular, the height and width are approximately equal to the compacton height and width. 
An approximate expression for highly supersonic solitary waves is obtained,
which gives a good approximation of the peak and the tail of these waves. 

Fifth, 
the profile of slightly supersonic solitary waves is shown to be approximately 
a sech-squared profile which is the same as solitons of the Korteweg-de Vries (KdV) equation. 
In addition, a two-scale asymptotic expansion of the LWHC equation around a fixed 
background strain is shown to yield the KdV equation with a scaled time variable 
and a scaled space variable in a reference frame moving with the sound speed. 

Finally, 
solitary waves with the same speed are compared across different nonlinearity regimes. 
This type of comparison has not been carried out previously. 

The paper is organized as follows. 

In section~\ref{model}, 
we give a full derivation of the LWHC wave equation from 
the equations of motion for a one-dimensional homogeneous chain of $N\gg 1$ discrete particles with arbitrary (non-zero) pre-compression. 
We discuss the total strain and background strain for solutions of the LWHC wave equation,
and show how the regimes of weak and strong nonlinearity can be formulated directly
in terms of these two strains.
We also discuss linearized wave solutions and derive their dispersion relation
from which the sound speed is obtained. 
Finally, we review the conservation laws for impulse momentum, energy, and momentum
in the LWHC wave equation. 

In section~\ref{solns},
we first review an exact quadrature expression for all solitary wave solutions. 
In the two cases $k=2$ and $k=3$,  
we note that the solitary waves have explicit expressions in terms of elementary functions,
which were derived in our previous work \cite{PrzAnc}. 
In the case $k=3/2$, 
we present an implicit expression for the solitary waves in terms of elliptic functions. 
This result has not appeared previously in the literature. 

In section~\ref{properties},
we give the expressions for the height, width, impulse and energy of the solitary waves. 
Using these expressions, we discuss physical properties of the solitary waves
and their dependence on the wave speed.  
We examine the properties in more detail in the cases when the wave speed is 
slightly supersonic and highly supersonic. 

In section~\ref{compare},
we explain how slightly supersonic waves have a weak nonlinearity,
and highly supersonic waves have a strong nonlinearity. 
In the strongly nonlinear case, 
we show that the solitary waves and the compacton have similar features,
and we present a useful approximate expression for the profile of the solitary waves. 
In the weakly nonlinear case, 
we show that the solitary waves are scaled KdV solitons. 
Finally, we present and discuss some scaling relations that hold among the energy, impulse, height and speed of solitary waves in both nonlinearity regimes. 

We make some concluding remarks in section~\ref{remarks}. 

Some analytical details of our results are given in three appendices.

\section{Long-wavelength continuum wave equation and its conservation laws}\label{model}

The equations of motion for a one-dimensional homogeneous chain of $N\gg 1$ discrete particles
interacting by a general power-law contact potential are given by 
\begin{equation}\label{discrete-eom}
m \ddot U_{i} = (k+1) a \big( (\delta_0-(U_{i}-U_{i-1}))^k - (\delta_0 -(U_{i+1}-U_{i}))^k \big),
\quad
i=2,\ldots, N-1
\end{equation}
in terms of the particle displacements $U_{j}(t)$, $j=1,\ldots,N$, 
relative to their initial (equilibrium) positions, 
where $\delta_0\geq 0$ is the initial overlap between adjacent particles 
due to pre-compression of the chain at $t=0$, 
and $k>1$ is determined by the geometry of their contact surface. 
Here $m$ is the particle mass, 
and $a$ is a constant which depends on the particles' material properties. 
The particles at each end of the chain obey similar equations of motion 
with a different potential that takes into account the boundary conditions. 
Since we will be interested in the continuum limit, in which $N\to\infty$, 
we will only need to consider the equations of motion \eqref{discrete-eom}
for the $N-2$ interior particles. 

In this system, the initial particle displacements are $U_{j}(0)=0$, $j=1,\ldots,N$ 
(since the chain is initially in its equilibrium configuration) at $t=0$. 
The system is initially uncompressed if $\delta_0=0$,
in which case the initial separation between the center of mass of adjacent particles is $2R$,
where $R$ is the particle radius. 
Instead, if $\delta_0>0$, the system has a pre-compression, 
in which case the initial separation between the center of mass of adjacent particles is $2R-\delta_0$.
For $t>0$, 
the dynamical overlap between adjacent particles is given by 
\begin{equation}\label{overlap}
\delta(t;i) = \begin{cases}
\delta_0+U_{i}(t)-U_{i+1}(t)  & \text{ if } U_{i+1}(t)-U_{i}(t)<  \delta_0\\
0 & \text{ if } U_{i+1}(t)-U_{i}(t)  \geq \delta_0 
\end{cases}
\end{equation}
where the property 
\begin{equation}\label{contact}
\delta(t;i)\geq 0 
\end{equation}
enforces the contact nature of the potential. 

In typical experiments and numerical simulations, 
the initial particle velocities are $\dot U_{j}(0)=0$ for the interior particles $j=2,\ldots,N-1$,
while for the two end particles, 
the velocities $\dot U_{1}(0)$ and $\dot U_{N}(0)$ 
correspond to imparting an initial sharp (short duration) impulse at one end or both ends of the chain.
Such an impulse gives the chain a fixed amount of energy $\tfrac{1}{2}m \big ( \dot U^2_{1}(0) + \dot U^2_{N}(0) \big )$
\cite{Nesterenko1985,Sokolow2007,Job2007}. 
In this situation, a propagating compressive pulse is produced at each end in the system, 
where the wave amplitude is described by the dynamical overlap variable \eqref{overlap},
with 
\begin{equation}\label{initialoverlap}
\delta(0;i)=\delta_0 
\end{equation}
being the initial (background) amplitude. 
Since the pulse is compressive, 
the moving particles in the pulse will satisfy 
$U_{i+1}-U_{i}<0$. 

Rarefactive pulses \cite{YanLiuYanZhaDuaYan,Yasuda2017,PrzAnc} 
can be produced by imparting more complicated initial conditions to the discrete system. 
The moving particles in these pulses will satisfy 
$0\leq U_{i+1}-U_{i} <\delta_0$. 
Note that $U_{i+1}-U_{i} =\delta_0$ corresponds to a broken contact between the particles. 

For any type of pulse, 
if the dynamical overlap is recorded for one (or a few) particle(s), 
it gives the wave amplitude $\delta(t;i)$ as a function of time $t$ 
at one (or a few) fixed value(s) of $i$. 
When travelling wave pulses are considered, 
the amplitude profile $\delta(t;i)$ in $t$ will have the same shape (up to scaling)
as a snapshot of the amplitude $\delta(t_0;i)$ 
as a function of particle number $i$ at a fixed time $t=t_0$. 

The dynamical state of a precompressed system is said to be \emph{strongly compressed}
if the relative displacement between adjacent moving particles is much smaller than the size of the initial overlap: 
\begin{equation}\label{strong:U}
|U_{i+1}(t)-U_{i}(t)|=|\delta(t;i)-\delta_0| \ll \delta_0, 
\quad
t>0.
\end{equation}
In this case the dynamical overlap will be approximately the same magnitude as the initial overlap, 
\begin{equation}\label{strong}
\delta(t;i)\approx \delta_0,
\end{equation}
and the resulting motion of the particles in the system will be \emph{weakly nonlinear}. 

Instead if the dynamical overlap is large compared to the initial overlap,
then the dynamical state of a precompressed system is said to be \emph{weakly compressed}:
\begin{equation}\label{weak}
\delta(t;i)\gg \delta_0,
\quad
t>0.
\end{equation}
In this case the relative displacement between adjacent moving particles will be at least the size of the initial overlap, 
\begin{equation}
U_{i}(t)-U_{i+1}(t) \gtrsim \delta_0.
\end{equation}
and the resulting motion of the particles in the system will be \emph{strongly nonlinear}. 

To derive a continuum wave equation for long wavelength pulses 
in a discrete system with arbitrary pre-compression $\delta_0\geq 0$, 
it is useful to begin with a change of variables 
\begin{equation}\label{displacement}
U_i = u_i +i\delta_0, 
\quad
i=1,\dots,N . 
\end{equation}
Physically, 
$u_i$ is the displacement of the $i$th particle as measured 
with respect to a reference system that has zero pre-compression. 
In terms of this variable, 
the equations of motion \eqref{discrete-eom} of the precompressed system become 
\begin{equation}\label{discrete-sys}
m \ddot u_{i} = (k+1) a \big( (u_{i-1} -u_{i})^k - (u_{i} -u_{i+1})^k \big),
\quad
i=2,\ldots, N-1.
\end{equation}
Then the continuum limit for this system \eqref{discrete-sys} is obtained by putting
\begin{equation}
u_{i}(t) \to u(t,x), 
\quad
u_{i\pm1}(t) \to u(t,x\pm 2R) = e^{\pm2R\partial_x}u(t,x), 
\quad
i=2,\dots,N-1,
\end{equation}
with 
$u_{i-1} -u_{i} \to W_-$ and $u_{i} -u_{i+1} \to W_+$, 
where
\begin{equation}
\pm W_\pm = u(t,x)  - u(t,x\pm2R)
= u - e^{\pm2R\partial_x}u 
= -\sum_{n=1}^{\infty} \tfrac{(\pm 2R)^{n}}{n!} \partial_x^{n} u.
\end{equation}
This yields 
\begin{equation}\label{CWE}
\tfrac{m}{a(k+1)} u_{tt} = W_-^k - W_+^k 
\end{equation}
which is a continuum wave equation for $u(t,x)$. 

The continuum limit of the wave amplitude \eqref{overlap} is given by 
\begin{equation}\label{amplitude}
\delta(t;i) = u_{i}-u_{i+1} \to W_+
\end{equation}
which is a nonlocal variable,
while the continuum limit of the displacement variable \eqref{displacement} 
is given by 
\begin{equation}\label{U}
U_i(t) = u_i(t)+i\delta_0 \to U(t,x) = u(t,x) + \tfrac{\delta_0}{2R}x . 
\end{equation}
It will be natural to suppose that the linear mass density of the discrete system is preserved in the continuum limit. 
This yields 
\begin{equation}\label{massdens}
\rho = m/(2R-\delta_0)
\end{equation}
for the continuum mass density. 

The main step now consists of making a long wavelength expansion of 
the continuum wave equation \eqref{CWE} in terms of $u(t,x)$. 
Consider a pulse of wavelength $\ell$, 
where $u_x$ is taken to be $O(1)$ while the derivative of $u_x$ is $O(1/\ell)$. 
The condition for the wavelength to be long compared to the particle size 
is that $\ell\gg 2R$. 
Then the nonlinear terms in the wave equation \eqref{CWE} 
can be expanded in terms of the parameter $\epsilon = 2R/\ell \ll 1$
by using the relation $(2R)^n\partial_x^n u_x = O(\epsilon^n)$. 
Thus, we have 
\begin{equation}
\begin{aligned}
W_\pm^k & = 
(2R)^k\Big( -u_x -\sum_{n=1}^{\infty} \tfrac{(\pm 2R)^{n}}{(n+1)!} \partial_x^{n} u_x \Big)^k \\
& = 
(2R)^k \Big( 
(-u_x)^k \pm k R (-u_x)^{k-1} (-u_{xx}) 
+kR^2 (-u_x)^{k-2}\big( \tfrac{2}{3}(-u_x)(-u_{xxx})
\\&\qquad
+\tfrac{1}{2}(k-1)(-u_{xx})^2 \big)
\pm kR^3 (-u_x)^{k-3}\big( \tfrac{1}{3}(-u_x)^2(-u_{xxxx})
\\&\qquad
+\tfrac{2}{3}(k-1)(-u_x)(-u_{xx})(-u_{xxx})
+\tfrac{1}{6}(k-1)(k-2)(-u_{xx})^3 \big)
+O(\epsilon^4) \Big) . 
\end{aligned}
\end{equation}
This expansion yields
\begin{equation}
\begin{aligned}
\tfrac{m}{a(k+1)} u_{tt} = W_-^k - W_+^k 
& = -k(2R)^{k+1} \Big( 
(-u_x)^{k-1} (-u_{xx}) 
+\tfrac{1}{6}R^2\big( (k-1)(k-2) (-u_x)^{k-3}(-u_{xx})^3 
\\&\qquad
+4(k-1)(-u_x)^{k-2} (-u_{xx})(-u_{xxx}) +2(-u_x)^{k-1} (-u_{xxxx}) \big)
+O(\epsilon^4) \Big).
\end{aligned}
\end{equation}
Truncating the expansion at this order produces 
the well-known highly nonlinear fourth-order wave equation 
\cite{Nesterenko1994,Porter2008,Porter2009,Nesterenko2001}
for $u(t,x)$:
\begin{equation}\label{LWCE-u}
c^{-2} u_{tt} = (-u_x)^{k-1} u_{xx}
+\alpha (-u_x)^{k-3}u_{xx}^3 -\beta (-u_x)^{k-2} u_{xx}u_{xxx} +\gamma (-u_x)^{k-1} u_{xxxx} 
\end{equation}
where
\begin{equation}\label{coeffs}
\alpha = \tfrac{1}{6}R^2(k-1)(k-2), 
\quad
\beta = \tfrac{2}{3}R^2(k-1), 
\quad
\gamma = \tfrac{1}{3}R^2
\end{equation}
and 
\begin{equation}\label{c}
c^2 = ak(k+1)(2R)^{k+1}/m. 
\end{equation}
Note, in these expressions \eqref{coeffs}--\eqref{c}, 
the constant $c$ has units of speed,
and the constants $\alpha$, $\beta$, $\gamma$ have units of length-squared,
while $u$ has units of length. 

We will call equation \eqref{LWCE-u} 
the \emph{long-wavelength Hertzian continuum} (LWHC) wave equation. 
We emphasize that it is a valid long-wavelength continuum limit of 
the discrete system \eqref{discrete-eom} 
with arbitrary pre-compression $\delta_0\geq 0$. 

In the same limit, 
the wave amplitude \eqref{amplitude} corresponds to the $O(1)$ term in the expansion 
$W_+ = (2R)\big( -u_x +O(\epsilon) \big)$. 
This term, given by $W_+ \approx -(2R)u_x$, 
is proportional to the dimensionless strain defined by 
\begin{equation}\label{strain}
v = -u_x . 
\end{equation}
Note this expression \eqref{strain} physically represents the \emph{total strain} (in the long-wavelength continuum limit) 
which includes a contribution from the precompression $\delta_0$. 
In particular, if we use the continuum limit of the particle displacement \eqref{U}
and define the corresponding strain variable $V=-U_x$, 
then we have the relation
\begin{equation}\label{dynstrain}
V = v-\tfrac{\delta_0}{2R}
\end{equation}
which has the physical meaning of the \emph{dynamical strain} (in the long-wavelength continuum limit). 

In terms of the total strain \eqref{strain},
we see that the condition \eqref{contact} enforcing the contact nature of the underlying potential is simply given by 
\begin{equation}\label{straincond}
v > 0 .
\end{equation}
Note that we exclude the possibility $v=0$ 
because it would correspond to having no contact between adjacent particles in the underlying discrete system. 
The initial precompression \eqref{initialoverlap} corresponds to the initial condition 
\begin{equation}\label{initialstrain}
v|_{t=0} =v_0= \tfrac{\delta_0}{2R} 
\end{equation}
which describes a background strain. 
On physical grounds, 
\begin{equation}\label{initialstrain_conds}
0< v_0 < 1 . 
\end{equation}

Compressive pulses in the continuum limit are described by the property 
$V\geq 0$ for the dynamical strain, 
whereas rarefactive pulses have the property $0\geq V > -v_0$. 
The weakly nonlinear regime 
(corresponding to strong compression \eqref{strong:U} in the discrete system)
is characterized by the dynamical condition 
$|V| \ll v_0$; 
the strongly nonlinear regime 
(corresponding to weak compression \eqref{weak} in the discrete system)
is characterized by the dynamical condition 
$V \gtrsim v_0$. 
These dynamical conditions can be expressed in terms of the total strain by 
\begin{equation}\label{weaknonlin:V}
|v- v_0| \ll v_0 
\end{equation}
for the weakly nonlinear regime, 
and 
\begin{equation}\label{strongnonlin:V}
v \gg v_0
\end{equation}
for the strongly nonlinear regime. 

Finally, we note that the total strain \eqref{strain} satisfies a long-wavelength wave equation 
given by the $x$-derivative of the LWHC wave equation:
\begin{equation}\label{LWCE-v}
c^{-2} v_{tt} = (v^{k-1} v_{x}+\alpha v^{k-3}v_{x}^3 +\beta v^{k-2} v_{x}v_{xx} +\gamma v^{k-1} v_{xxx})_x
\end{equation}
with $v$ satisfying the conditions \eqref{straincond} and \eqref{initialstrain}.

\subsection{Linearized (sound) waves}

In the continuum limit,  
waves that have very small amplitude relative to the size of the pre-compression 
physically describe the linearized sound waves of the continuum system. 
Long wavelength sound waves satisfy the linearized approximation of the long-wavelength wave equation \eqref{LWCE-v}, 
which is given in terms of the dynamical strain \eqref{dynstrain} by 
\begin{equation}\label{LWCE-V}
c^{-2} V_{tt} = (v_0^{k-1} V_{x}+\gamma v_0^{k-1} V_{xxx})_x
\end{equation}
where $\gamma=\tfrac{1}{3}R^2$. 
In this approximation, $V$ is small compared to the initial total strain $v_0$:
\begin{equation}
0<V\ll v_0=\tfrac{\delta_0}{2R} . 
\end{equation}
To obtain the dispersion relation, 
we substitute a harmonic mode expression $V=e^{i(\kappa x -\omega t)}$
into this wave equation, 
yielding $c^{-2} \omega^2 =v_0^{k-1} \kappa^2 -\tfrac{1}{3}R^2 v_0^{k-1} \kappa^4$.
Since the wavelength is $\ell = 2\pi/\kappa$,
which obeys $2R/\ell \ll 1$, 
we see that 
$c^{-2} \omega^2 =v_0^{k-1} (2\pi/\ell)^2( 1 -\tfrac{\pi^2}{3}(2R/\ell)^2)$
can be approximated by
$c^{-2} \omega^2 \simeq v_0^{k-1} (2\pi/\ell)^2$. 
This yields a linear dispersion relation 
\begin{equation}
\pm \omega \simeq v_0^{(k-1)/2} c\kappa. 
\end{equation}
Hence the sound speed is given by 
\begin{equation}\label{soundspeed}
c_0 = \omega/\kappa = \big(\tfrac{\delta_0}{2R}\big)^{(k-1)/2} c.
\end{equation}

\subsection{Conservation laws}

A conservation law for the LWHC wave equation \eqref{LWCE-u} 
is a local continuity equation 
\begin{equation}\label{conslaw}
D_t T + D_x X =0
\end{equation}
holding for all solutions $u(t,x)$ of the wave equation, 
where $T$ is the conserved density and $X$ is the spatial flux,
which are given by functions of $t$, $x$, $u$, $u_t$, and $x$-derivatives of $u$ and $u_t$. 
As shown in Ref.~\cite{PrzAnc}, 
all conservation laws with conserved densities of the first-order form $T(t,x,u,u_t,u_x)$ 
are given by a linear combination of the following four conservation laws:
\begin{align}
& \begin{aligned}
& T_1= -u_t u_x,
\\
& X_1 = \tfrac{1}{2} u_t^2 -c^2  \big( \tfrac{1}{3}R^2 (-u_x)^{k} u_{xxx} -\tfrac{k-2}{6} R^2 (-u_x)^{k-1} u_{xx}^2 -\tfrac{1}{k+1} (-u_x)^{k+1} \big); 
\end{aligned}
\label{TX1}\\
& \begin{aligned}
& T_2 = \tfrac{1}{2} u_t^2 - c^2 \big( \tfrac{1}{6}R^2 (-u_x)^{k-1} u_{xx}^2 - \tfrac{1}{k(k+1)} (-u_x)^{k+1} \big), 
\\
& X_2 = -c^2\big( \tfrac{1}{3}R^2 (-u_x)^{k-1}(u_t u_{xxx} -u_{tx} u_{xx})  -\tfrac{k-1}{6}R^2 (-u_x)^{k-2}u_t u_{xx}^2 - \tfrac{1}{k} (-u_x)^k u_t \big) ;
\end{aligned}
\label{TX2}\\
&\begin{aligned}
& T_3 = u_t,
\\
& X_3 = c^2\big( -\tfrac{1}{3}R^2 (-u_x)^{k-1} u_{xxx} +\tfrac{k-1}{6}R^2 (-u_x)^{k-2} u_{xx}^2 +\tfrac{1}{k} (-u_x)^k \big); 
\end{aligned}
\label{TX3}
\\
&\begin{aligned}
& T_4 = u -tu_t, 
\\
& X_4 = c^2 t \big( \tfrac{1}{3}R^2 (-u_x)^{k-1} u_{xxx} - \tfrac{k-1}{6}R^2 (-u_x)^{k-2} u_{xx}^2 -\tfrac{1}{k} (-u_x)^k \big) . 
\end{aligned}
\label{TX4}
\end{align}
In each of these expressions, 
we have omitted an overall factor consisting of the continuum mass density \eqref{massdens}. 
When this factor is restored, 
the conservation laws \eqref{TX1} and \eqref{TX2} respectively describe 
the total momentum and energy of solutions $u(t,x)$;
the conservation law \eqref{TX3} describes the total impulse,
while the conservation law \eqref{TX4} is connected with the mean value of $u(t,x)$.

\section{Exact solitary waves}\label{solns}

We are interested in travelling wave solutions 
\begin{equation}\label{u-travelwave}
u= \f(\zeta), 
\quad
\zeta = x-\V t
\end{equation} 
of the LWHC wave equation \eqref{LWCE-u}. 
Substitution of expression \eqref{u-travelwave}
into equation \eqref{LWCE-u} yields a fourth-order differential equation 
\begin{equation}\label{f-ode}
(\V/c)^2 \f'' = (-\f')^{k-1} \f'' + \alpha (-\f')^{k-3}(\f'')^3 -\beta (\f')^{k-2}\f''\f''' +\gamma (-\f')^{k-1} \f''''
\end{equation}
where $\V$ is the (constant) wave velocity,
and where $c$, $\alpha$, $\beta$, $\gamma$ are given by expressions \eqref{coeffs}--\eqref{c}. 

The physical variable that will support solitary waves is the dynamical strain \eqref{dynstrain}. 
For a travelling wave \eqref{u-travelwave}, this variable has the form 
\begin{equation}
V = -\f'(\zeta) -v_0  .
\end{equation}
Since a solitary wave has a localized profile in $\zeta$, 
we want solutions for which 
\begin{equation}
\lim_{\zeta\to\pm\infty} V =0 . 
\end{equation}
In terms of the total strain \eqref{strain}, 
solitary waves are described by 
\begin{equation}\label{f-cond}
v = -\f'(\zeta) >0
\end{equation}
and 
\begin{equation}\label{f-bc}
\lim_{\zeta\to\pm\infty} v = v_0 >0 . 
\end{equation}
Here $v_0$ is related to the pre-compression through the initial condition \eqref{initialstrain} 
and represents the asymptotic (background) value of $v$. 

It will be useful, mathematically, to work with dimensionless scaled variables:
\begin{equation}\label{scal-g-xi}
g = -\f'/\lambda, 
\quad
\xi =\zeta/l
\end{equation}
with 
\begin{equation}\label{scaling}
\lambda = \left(\tfrac{1}{2}k(k+1)(\V/c)^2 \right)^\frac{1}{k-1} ,
\quad
l = \sqrt{\tfrac{1}{6}k(k+1)}R .
\end{equation}
The inverse transformation from $g(\xi)$ to $v(t,x)$ is given by 
\begin{equation}\label{change_vars}
v(t,x)= \lambda g((x-\V t)/l).
%\big(\tfrac{1}{2} (\V/c)^2 k(k+1)\big)^{1/(k-1)} g\big((\sqrt{6}/(R\sqrt{k(k+1)}))(x-\V t)\big) 
\end{equation}

In terms of these scaled variables \eqref{scal-g-xi}, 
the travelling wave equation \eqref{f-ode} becomes 
a third-order differential equation 
\begin{equation}\label{g-ode}
0= 2 g^{k-1} g''' +4(k-1) g^{k-2}g'g'' +(k-1)(k-2)g^{k-3}g'{}^3 +k(k+1) g^{k-1} g' -2 g',
\end{equation}
and the asymptotic condition \eqref{f-bc} on solutions $g(\xi)$ becomes 
\begin{equation}\label{g-bc}
\lim_{\xi\to\pm\infty} g(\xi) = g_0 >0
\end{equation}
where
\begin{equation}\label{g0}
g_0 = v_0/\lambda = \tfrac{\delta_0}{2R} \left(\tfrac{1}{2}k(k+1)(\V/c)^2 \right)^\frac{1}{1-k} .
\end{equation}

The third-order differential equation \eqref{g-ode} can be directly reduced to 
a first-order separable differential equation 
by using first integrals that arise from the conservation laws \eqref{TX1}--\eqref{TX3} 
(which do not contain $t$ and $x$ explicitly)
admitted by the LWHC wave equation \eqref{LWCE-u}, 
as explained in Ref.~\cite{PrzAnc}. 
This yields the ODE 
\begin{equation}\label{scal-ode}
(g')^2 =  g^{1-k} (g^2 + C_1 g +C_2 - g^{1+k})
\end{equation}
where $C_1$ and $C_2$ are dimensionless arbitrary constants
which correspond to scaled first integrals. 
The physical meaning of the two first integrals are, respectively, 
the spatial flux of the impulse in the rest frame of the travelling wave, 
and the spatial flux of energy/momentum in the rest frame of the travelling wave. 
Thus, for any solution $g(\xi)$, 
$C_2=E$ represents a dimensionless energy 
and $C_1 =I$ represents a dimensionless impulse. 

We now view the first-order differential equation \eqref{scal-ode} 
as being analogous to the energy integral for motion in a potential well
\begin{equation}\label{ode_energy_form}
g^{k-1} g'{}^2 + \mathcal{V}(g) = E 
\end{equation}
where $g^{k-1} g'{}^2$ plays the role of the kinetic energy,
and where the potential energy is given by 
\begin{equation}\label{V}
\mathcal{V}(g) =  g^{1+k} -g(g+I) 
\end{equation}
which depends on a free parameter $I$. 
A comprehensive qualitative analysis has been carried out in Ref.~\cite{PrzAnc}
to determine the values of $(E,I)$ that lead to solitary wave solutions for $g(\xi)$
with the asymptotic boundary condition \eqref{g-bc},
and also to single-arch travelling wave (compacton) solutions for $g(\xi)$. 
This analysis, which we will summarize next, depends crucially on the shape of the potential \eqref{V}.

\subsection{Quadrature formula for solitary waves}\label{sec:solitary-examples}

Solitary waves correspond to motion for $g$ in which there is 
one turning point $g=g_1$ and one asymptotic equilibrium point $g=g_0$. 
These two points are roots of the effective energy equation $\mathcal{V}(g)=E$ 
such that $\mathcal{V}'(g_0)=0$ and $\mathcal{V}'(g_1)>0$. 
The condition of positive pre-compression \eqref{f-bc} 
combined with the shape of the potential $\mathcal{V}(g)$ 
implies that $g_1$ and $g_0$ belong to the intervals 
\begin{equation}\label{solitary_g0}
0<g_0<g^* < g_1< 1 
\end{equation}
where
\begin{equation}\label{gstar}
g^* = \big( \tfrac{2}{k(k+1)}\big)^\frac{1}{k-1} 
\end{equation}
is the inflection point of the potential, $V''(g^*)=0$. 
The relationship between the asymptotic equilibrium point $g_0$ and the parameters $(E,I)$ 
is given by 
\begin{equation}\label{solitary_EI}
E = g_0^2(1- k g_0^{k-1}), 
\quad
I = (k+1)g_0^k - 2g_0 
\end{equation}
which leads to the corresponding parameter ranges 
\begin{equation}\label{solitary_EI_range}
E>0,
\quad
0>I>I^*
\end{equation}
where 
\begin{equation}\label{Istar}
I^* = (1 + k) g^*{}^{k} - 2g^* = -\tfrac{2(k-1)}{k}g^* <0 . 
\end{equation}
Note $g_1$ is a function of $g_0$ 
as given by the positive root of the algebraic equation
\begin{equation}\label{g1}
0=\mathcal{V}(g_1)-E= g_1^{k+1} -g_1^2 +g_0(2-(k+1)g_0^{k-1})g_1 + g_0^2(kg_0^{k-1}-1)
\end{equation}
in the interval \eqref{solitary_g0}. 
As shown in \Ref{PrzAnc}, 
$g_1$ is a decreasing function of $g_0$,
with $g_1\to 1$ when $g_0\to 0$, 
and $g_1\to g^*$ when $g_0\to g^*$. 

It will be useful for mathematical purposes to note that 
$E-\mathcal{V}(g) = (g-g_0)^2 A(g,g_0)$ holds,
where 
\begin{equation}\label{A}
A(g,g_0)=1- \partial_{g_0}\Big(g_0 \frac{g^k-g_0^k}{g-g_0} \Big),
\end{equation}
which has the properties 
\begin{equation}\label{A_props}
A(g_0,g_0)= 1-(g_0/g^*)^{k-1},
\quad
A(g_1,g_0)=0 .
\end{equation}

All solitary wave solutions $g(\xi)$ are then given by the quadrature of the ODE \eqref{ode_energy_form}:
\begin{equation}\label{solitary_ode_integral}
\int_{g}^{g_1} \frac{\sqrt{g^{k-1}}}{(g-g_0)\sqrt{A(g,g_0)}} \,dg = |\xi| ,
\end{equation}
where $g_1$ is determined by the algebraic equation \eqref{g1} in terms of $g_0$,
and where $g_0$ obeys the inequality \eqref{solitary_g0}.
From this equation \eqref{solitary_ode_integral}, 
$g(\xi)$ has the following two main features. 

First, $g(\xi)$ has a single peak $g=g_1$ at $\xi=0$
and an asymptotic tail with $g\to g_0$ as $|\xi|\to \infty$. 
To see why, note that the extrema of $g(\xi)$ are determined by the roots of 
$0=E-\mathcal{V}(g)= (g-g_0)^2 A(g,g_0)$ in the interval $g_0\leq g\leq g_1$. 
The roots consist of $g=g_0$ and $g=g_1$. 
From the ODE \eqref{ode_energy_form}, note that 
$g''=-\tfrac{1}{2}\big( g\mathcal{V}'(g) +(k-1)(E-\mathcal{V}(g)) \big)g^{-k}$. 
The properties of the potential $\mathcal{V}(g)$ then show 
$g''<0$ when $g=g_1$ and $g''=0$ when $g=g_0$. 
Hence, $g_1$ represents the peak value of $g(\xi)$. 
When $g=g_0$, 
it is straightforward to see that $g''=0$, $g'''=0$, and so on. 
Moreover, the integral \eqref{solitary_ode_integral} clearly diverges as $g\to g_0$,
whereby $|\xi|\to \infty$. 
Hence, $g_0$ represents the asymptotic value of $g(\xi)$.  

Second, the asymptotic tail of $g(\xi)$ exhibits exponential decay 
$g\sim g_0 + (g_1-g_0)e^{\xi_0/\chi} e^{-|\xi|/\chi}$
for $|\xi|\gg \xi_0 +\chi$,
where the dimensionless scale for the decay is given by 
\begin{equation}\label{decayscale_g0}
\chi = 1/\sqrt{g_0^{1-k} - {g^*}^{1-k}} ,
\end{equation}
and the offset in $\xi$ is given by 
\begin{equation}\label{decayoffset_g0}
\xi_0 = \int_{g_0}^{g_1} \frac{F(g,g_0)}{g-g_0} \,dg 
\end{equation}
with 
\begin{equation}\label{F}
F(g,g_0) = 
\frac{g^\frac{k-1}{2}}{\sqrt{A(g,g_0)}} - \frac{g_0{}^\frac{k-1}{2}}{\sqrt{A(g_0,g_0)}} . \end{equation}
To derive this asymptotic behaviour, 
we split up the integral \eqref{solitary_ode_integral} 
by extracting the leading-order term that involves $g$, 
a subleading term that is constant, and a remainder term that vanishes when $g=g_0$:
\[
\int_{g}^{g_1} \frac{g^\frac{k-1}{2}}{(g-g_0)\sqrt{A(g,g_0)}} \,dg 
= \frac{g_0{}^\frac{k-1}{2}}{\sqrt{A(g_0,g_0)}} \int_{g}^{g_1} \frac{dg}{g-g_0}
+\int_{g_0}^{g_1} \frac{F(g,g_0)}{g-g_0} \,dg 
-\int_{g_0}^{g} \frac{F(g,g_0)}{g-g_0} \,dg .
\]
The first and second integrals on the righthand side yield, respectively, 
$\chi\ln\big((g_1-g_0)/(g-g_0)\big)$ and $\xi_0$,
while the remaining integral is $O(g-g_0)$ because $F(g,g_0)/(g-g_0)= O(1)$
due to $F(g_0,g_0)=0$. 
This gives 
\[
\int_{g}^{g_1} \frac{g^\frac{k-1}{2}}{(g-g_0)\sqrt{A(g,g_0)}} \,dg 
\simeq \chi\ln\Big(\frac{g_1-g_0}{g-g_0}\Big) + \xi_0 
\]
for $g\simeq g_0$.
The quadrature \eqref{solitary_ode_integral} then yields 
$\xi_0 -|\xi| \simeq \chi\ln\big((g-g_0)/(g_1-g_0)\big)$,
which establishes the leading-order term in the asymptotic form for $g-g_0$.

\subsection{Explicit solitary wave expressions}

We showed in Ref. \cite{PrzAnc} that the quadrature \eqref{solitary_ode_integral}
can be evaluated in terms of elementary functions when $k=2,3$
and in terms of elliptic functions when $k=\tfrac{3}{2},4,5$. 
All of these cases lead to implicit algebraic expressions for $g(\xi)$. 

\begin{figure}[!h]
\centering
\includegraphics[width=\textwidth]{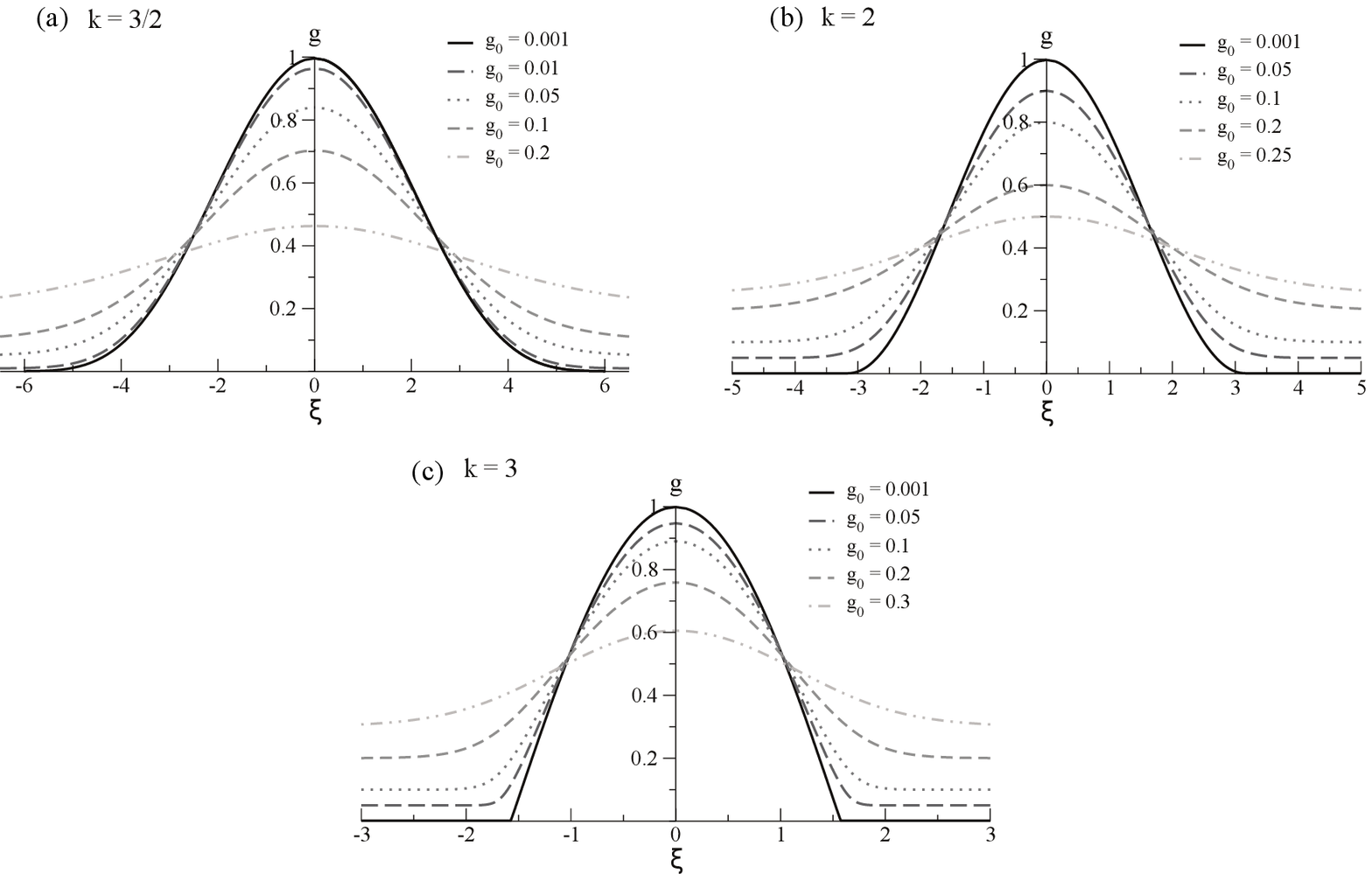}
\caption{
Solitary wave solutions \eqref{solitarysol3}, \eqref{solitarysol1} and \eqref{solitarysol2}, for $k=\tfrac{3}{2}, 2$ and $3$, respectively.
}
\label{fig:solution_plots}
\end{figure}

For the cases $k=2$ and $k=3$, 
the respective solutions $g(\xi)$ are given by 
\begin{equation}
\frac{g_0}{\sqrt{P(g_0)}} \arctanh\Big(Q(g)\Big) 
+\arctan\bigg( \frac{\sqrt{P(g)}}{g + g_0-\tfrac{1}{2}} \bigg) 
=|\xi|,
\label{solitarysol1}
\end{equation}
with 
\begin{equation}
P(g) = (g_1-g)g, 
\quad
g_1 = 1-2g_0,
\quad
g^*=1/3,
\label{solitarysol1-P}
\end{equation}
and
\begin{equation}
\frac{g_0}{\sqrt{P(g_0)}} \arctanh\Big(Q(g)\Big) 
+\arctan\bigg( \frac{\sqrt{P(g)}}{g + g_0} \bigg) 
=|\xi| ,
\label{solitarysol2}
\end{equation}
with 
\begin{equation}
P(g) = (g_1-g)(g+g_1+2g_0),
\quad
g_1 = \sqrt{1-2g_0^2}-g_0,
\quad
g^*=1/\sqrt{6},
\label{solitarysol2-P}
\end{equation}
where 
\begin{equation}\label{Q}
Q(g)=\frac{2\sqrt{P(g)P(g_0)}}{P(g)+ P(g_0)+(g-g_0)^2} . 
\end{equation}
In both cases, 
$g_0$ is a free parameter in the range $0<g_0<g^*$, 
and $g_1$ is given explicitly in terms of $g_0$.
(Note, here $\arctan$ is defined to be continuous in the given range for $g_0$.)
See Fig.~\ref{fig:solution_plots}(b,c) for the solution profiles for different $g_0$.

For the case $k=\tfrac{3}{2}$, 
the solution $g(\xi)$ is given by 
\begin{equation}\label{solitarysol3}
I(\sqrt{g}) + J(\sqrt{g}) = |\xi|,
\quad
g=h^2
\end{equation}
where
\begin{equation}\label{solitarysol3-rat-terms}
\begin{aligned}
I(h) & = 
\sigma \ln\bigg(\frac{2\sqrt{\Gamma}\big(Y(h)/Y_0+\sqrt{X(h)/X_0}\big)\big(\sqrt{X(h)}+\sqrt{X_0}\big)((h+h_2)^2+h_3^2)}{\Lambda(h-h_0)(1+Y(h)/Y_0)\sqrt{1+\varpi X(h)}}\bigg)
\\&\qquad
+ \arctan\bigg( \frac{2\sqrt{X(h)}}{1-X(h)} \bigg)
\end{aligned}
\end{equation}
is a sum of elementary functions given in terms of the rational functions
\begin{equation}
X(h)=\frac{h(h_1-h)}{(h+h_2)^2+h_3^2},
\quad
Y(h)=\frac{h-h_1/\Phi_+}{h +h_1/\Phi_-}
\end{equation}
and where 
\begin{equation}\label{solitarysol3-ellipt-terms}
\begin{aligned}
J(h) & = \eta \big( (\theta-\phi)Z(h) + \Pi\big({-\mu};Z(h)|\psi\big) -\theta \Pi\big(\nu;Z(h)|\psi\big) \big)
\end{aligned}
\end{equation}
is a sum of Jacobi elliptic functions (of the third kind)~\cite{Lawden1980}
\begin{equation}
\begin{aligned}
& 
\Pi(n;\theta|l) 
%= \int_{0}^{\sn(\theta|l)} \frac{dz}{(1-nz^2)\sqrt{(1-z^2)(1-lz^2)}}
= \frac{1}{\sqrt{1+j}} \int_{\cn(\theta|l)}^{1} \frac{dz}{(mz^2-1)\sqrt{(1-z^2)(1+jz^2)}},
\\
& l=j/(j+1), 
\quad
n =m/(m-1), 
\quad
j>0, 
\end{aligned}
\end{equation}
with 
\begin{equation}\label{solitary-sol3-cn_inverse}
Z(h)= \cn^{-1}\big((\Phi_+/\Phi_-)Y(h)|\psi\big)
\end{equation}
being given by the inverse of the $\cn$ elliptic function. 
The constants in expressions \eqref{solitarysol3-rat-terms}--\eqref{solitary-sol3-cn_inverse} 
are given by 
\begin{gather}
X_0= X(h_0), 
\quad
Y_0= Y(h_0),
\quad
W_0 =(h_0+h_2)^2 +h_3^2,
\quad
W_1 =(h_1+h_2)^2 +h_3^2,
\label{solitarysol3-const1}
\\
\sigma =\frac{2h_0^2}{\sqrt{h_0(h_1-h_0)W_0}},
%=\frac{2h_0^2}{\sqrt{h_0(h_1-h_0)B_2(h_0)}}
\quad
\varpi 
%= \frac{(\Phi_+h_0 -h_1)^2(1-B_2(0)\Phi_-^2/h_1^2)}{(\Phi_- h_0+h_1)^2}
%= \frac{Y(h_0)^2}{Y(h_1)^2}(1-B_2(0)\Phi_-^2/h_1^2)
= \frac{Y_0^2\Phi_+^2}{\Phi_-^2}(1-1/\Omega^2),
\label{solitarysol3-const2}
\\
\eta 
%= 2\frac{\Phi_+h_1}{2\Phi_-\sqrt[4]{B_2(0)B_2(h_1)}}
= \frac{2\Omega\Phi_+}{\Phi_-\sqrt{2(\Phi_++\Phi_-)}},
\quad
\phi 
%= \frac{(1-Y(h_0))\sqrt{B_2(0)}}{\sqrt{B_2(h_0)}}
= (1-Y_0)\Lambda,
\quad
\theta = \frac{h_0 Y_0}{h_1-h_0},
\label{solitarysol3-const3}
\\
\psi
%= \frac{h_1^2-B_2(0)\Phi_-^2}{4\sqrt{B_2(0)B_2(h_1)}},
= (\Omega^2-1)\frac{\Phi_-^2}{2(\Phi_++\Phi_-)},
\quad
\mu
%= \frac{\Phi_-^2\sqrt{B_2(0)}}{4\sqrt{B_2(h_1)}}
%= \frac{\Phi_-^2}{2(\Phi_++\Phi_-)}
= \frac{\psi}{\Omega^2-1},
\quad
\nu 
%= \frac{B_1(h_0)(h_1^2/B_2(0) -\Phi_-^2)}{(\Phi_- h_0+h_1)^2}
%= (\Omega^2-1)\Phi_-^2\Gamma
=\frac{\psi\Gamma}{2(\Phi_+ +\Phi_-)},
\label{solitarysol3-const4}
\\
\Phi_\pm = \frac{\sqrt{W_1}}{\sqrt{h_2^2+h_3^2}} \pm 1,
%= \frac{\sqrt{B_2(h_1)}}{\sqrt{h_2^2+h_3^2}} \pm 1
%= \frac{\sqrt{B_2(h_1)}}{\sqrt{B_2(0)}} \pm 1
\quad
\Omega = \frac{h_1}{\Phi_-\sqrt{h_2^2+h_3^2}},
%= \frac{h_1}{\Phi_-\sqrt{B_2(0)}}
\quad
\Gamma = \frac{h_0(h_1-h_0)}{(\Phi_-h_0+h_1)^2},
\quad
\Lambda = \frac{\sqrt{h_2^2+h_3^2}}{\sqrt{W_0}}, 
\label{solitarysol3-const5}
\end{gather}
with
\begin{align}
& h_1 = \tfrac{1}{3}-\tfrac{2}{3}h_0 + r_-^{1/3} - r_+^{1/3}, 
\label{h1-expr}\\
& h_2 = -\tfrac{1}{3}+\tfrac{2}{3}h_0 + \tfrac{1}{2}(r_-^{1/3} - r_+^{1/3}), 
\label{h2-expr}\\
& h_2^2+h_3^2 = (h_0 -\tfrac{2}{3})h_0 + r_-^{2/3} +r_+^{2/3} + (\tfrac{2}{3}h_0-\tfrac{1}{3})( r_-^{1/3} - r_+^{1/3} ),
\label{h2h3-expr}\\
& r_{\pm} = 
\tfrac{5}{12} \sqrt{h_0^3(h_0-\tfrac{2}{3})(h_0^2-\tfrac{2}{5}h_0-\tfrac{4}{25})} \pm \tfrac{5}{108}(h_0^3+\tfrac{24}{5}h_0^2-\tfrac{12}{5}h_0-\tfrac{4}{5}) .
\end{align}
In this solution, 
all of these constants depend only on $h_0$ 
which is a free parameter in the range $0<h_0<h^*=\tfrac{8}{15}$,
corresponding to $0<g_0<g^*=\tfrac{64}{225}$. 
Furthermore, the constants \eqref{solitarysol3-const1}--\eqref{solitarysol3-const5} are positive. 
(Some details of the derivation of the solution are shown in Appendix~\ref{appendix:k=3/2}.)
See Fig.~\ref{fig:solution_plots}(a) for an illustration of the solution profile for different values of $g_0$.

The solutions for the other two cases $k=4$ and $k=5$ are similar to the case $k=\tfrac{3}{2}$ 
and will be omitted. 

\subsection{\label{sec:compacton}Single-arch travelling wave (compacton)}

The compacton corresponds to motion for $g$ in which there is 
one turning point $g=g_1=1$ and one equilibrium point $g=g_0=0$, 
with $\mathcal{V}'(0)=\mathcal{V}(0)=0$ and $\mathcal{V}'(1)>0$. 
The parameters $(E,I)$ are given by 
\begin{equation}\label{compacton_EI}
E = I = 0 
\end{equation}
which are the limiting values of $E$ and $I$ in the case of solitary waves 
when $g_0$ approaches $0$. 
%We note that the potential \eqref{V} with $I=0$ no longer has a local maximum for $g>0$. 
For these values \eqref{compacton_EI}, 
the solution $g(\xi)$ of the ODE \eqref{ode_energy_form} is given by the quadrature
\begin{equation}\label{nodal_ode_integral}
\int_{g}^{1} \frac{\sqrt{g^{k-3}}}{\sqrt{1-g^{k-1}}} \,dg = |\xi| . 
\end{equation}
Here the integral can be evaluated explicitly for all $k>1$, 
yielding
\begin{equation}\label{periodicsol}
g(\xi) = \cos\big(\tfrac{1}{2}(k-1)|\xi-L\lfloor \tfrac{1}{2} + \xi/L\rfloor|\big)^\frac{2}{k-1}\geq 0, 
\quad 
L=\tfrac{2}{k-1}\pi
\end{equation}
where $\lfloor x \rfloor$ denotes the floor function. 
The solution \eqref{periodicsol} is a periodic function of $\xi$ 
having a peak $g=1$ at $\xi=0 \mod L$,
and a node $g=0$ at $\xi=\pm L/2 \mod L$. 
The node is a minimum if $1<k<3$, or a corner if $k=3$, or a cusp if $k>3$. 
We remark that this solution \eqref{periodicsol}
has appeared in \Ref{Nesterenko2001,Porter2009} 
but with the phase shift omitted. 
Without the phase shift, the power $\tfrac{2}{k-1}$ must be interpreted as 
the $(k-1)^\th$ positive real root of the square of the cosine. 

Since the solution \eqref{periodicsol} has a node at $\xi=\pm L/2$, 
we can cut off the solution expression at these two points 
and take $g(\xi)$ to vanish outside the domain $-L/2\leq \xi \leq L/2$. 
This yields a piecewise expression 
\begin{equation}\label{compactonsol}
g(\xi) =
\begin{cases}
\cos\big(\tfrac{1}{2}(k-1)\xi\big)^\frac{2}{k-1}, & |\xi|\leq \tfrac{1}{k-1}\pi
\\
0 , & |\xi|\geq \tfrac{1}{k-1}\pi
\end{cases}
\end{equation}
which is continuous at the nodal points $\xi=\pm L/2$. 
However, 
this piecewise expression needs to be three-times differentiable for it to be an actual solution of the travelling wave ODE \eqref{g-ode}. 
In particular, we must have 
$0=g_\per(\pm L/2)=g_\per'(\pm L/2)=g_\per''(\pm L/2)=g_\per'''(\pm L/2)$,
where $g_\per(\xi)$ denotes the periodic solution \eqref{periodicsol}. 
These conditions are easily verified to hold if and only if $1<k<\tfrac{5}{3}$. 
More specifically, at the cutoff $\xi=\pm L/2$, 
we see $g_\per'(\xi)$ is discontinuous when $k\geq 3$; 
$g_\per''(\xi)$ is discontinuous when $k\geq 2$; 
and $g_\per'''(\xi)$ is discontinuous when $k\geq \tfrac{5}{3}$. 

The resulting solution \eqref{compactonsol}, 
which consists of a single arch of the periodic solution \eqref{periodicsol}, 
is a compacton. 
It exists only for $k$ in the range 
\begin{equation}
1<k<\tfrac{5}{3} .
\end{equation}
This condition on $k$ is often overlooked in the literature.

\section{Physical properties of the solitary waves}\label{properties}

We first express the quadrature formula \eqref{solitary_ode_integral}
for solitary waves in terms of the physical variables \eqref{change_vars} 
representing the total strain \eqref{strain}. 

All solitary waves $v=v(x-\V t)$ are determined implicitly by 
\begin{equation}\label{phys_solitarywave}
\frac{c}{\sqrt{3}|\V|}\int_{v}^{v_1} \frac{\sqrt{v^{k-1}}}{(v-v_0)\sqrt{\hat A(v,v_0;|\V|/c)}} \,dv = \frac{|x-\V t|}{R}
\end{equation}
where $\hat A(v,v_0;|\V|/c)=A(v/\lambda,v_0/\lambda)$ is given by the function \eqref{A}, 
and $v_1$ is the positive root of the algebraic equation 
\begin{equation}\label{v1Aeqn}
\hat A(v_1,v_0;|\V|/c)=0 , 
\end{equation}
such that 
\begin{equation}\label{v0v1interval}
0<v_0 < v^* < v_1 < \lambda
\end{equation}
with
\begin{equation}\label{vstar}
v^* = (|\V|/c)^\frac{2}{k-1} =  \big(\tfrac{2}{k(k+1)}\big)^\frac{1}{k-1}\lambda .
\end{equation}

The shape of the solitary waves consists of a single peak $v=v_1$ at $x=\V t$
and an asymptotic tail $v\to v_0$ as $|x|\to\infty$,
with the value of $v_0$ determined by the initial (background) pre-compression \eqref{initialstrain}. 
For $v_0\to 0$, we have $v_1\to \lambda$,
while for $v_0\to v^*$, we have $v_1\to v^*$. 

A useful observation is that the background strain $v_0$ is related to the sound speed $c_0$
through equations \eqref{initialstrain} and \eqref{soundspeed}:
\begin{equation}\label{v0c0rel}
c_0/c = v_0{}^\frac{k-1}{2} . 
\end{equation}  
As a consequence, 
the inequality $v_0<v^*$ 
can be expressed as the corresponding physical relation 
\begin{equation}\label{minspeed}
|\V| > c_0
\end{equation}  
between the solitary wave speed $\V$ and the sound speed $c_0$. 
This shows that all solitary waves are supersonic, 
which is a well-known statement in the literature 
in the case of weak compression. 
Here we see that it holds for arbitrary compression.

\subsection{Height and width}

Solitary waves describe physical compression waves measured by the total strain,
where $v_1$ is the maximum (peak) strain of the wave
and $v_0$ is the asymptotic background strain
on which the wave is propagating. 
The amplitude $v$ of a solitary wave away from the peak exhibits a spatial decay to the background amplitude $v_0$. 
As shown by the results in Section~\ref{solns}, 
this decay is exponential 
\begin{equation}\label{expdecay}
v-v_0 \sim 
%(v_1-v_0) e^{-(|x-\V t|-\zeta_0)/l_\s}
(v_1-v_0)e^{\zeta_0/l_\s} e^{-|x-\V t|/l_\s},
\quad
|x-\V t| \gg \zeta_0 + l_\s
\end{equation}
with respect to the length scale $l_\s=l\chi$
and the offset $\zeta_0 = \xi_0 l$,
given in terms of expressions \eqref{scaling}, \eqref{decayscale_g0}, \eqref{decayoffset_g0}. 
Explicit expressions for the length scale and the offset are given by 
\begin{equation}\label{decayscale}
l_\s =\tfrac{1}{\sqrt{3}} R\big/\sqrt{(\V^2/c^2) v_0^{1-k} - 1}
\end{equation}
and
\begin{equation}\label{decayoffset}
\zeta_0 = \sqrt{\tfrac{k(k+1)}{6}}R\int_{v_0}^{v_1} \frac{\hat F(v,v_0;|\V|/c)}{v-v_0} \,dv 
\end{equation}
where $\hat F(v,v_0;|\V|/c) = F(v/\lambda,v_0/\lambda)$ is given by the function \eqref{F}. 

In terms of the dynamical strain \eqref{dynstrain}, 
solitary waves have a peak amplitude 
\begin{equation}\label{height}
h_\s = v_1 - v_0 
\end{equation}
which defines the height of the solitary wave. 
For a fixed wave speed $\V$, 
this height has the properties 
$h_\s\to 0$ as $v_0\to v^*$,
and 
$h_\s \to \lambda$ as $v_0\to 0$,
where
$\lambda = (\tfrac{k(k+1)}{2}\V^2/c^2)^\frac{1}{k-1}$. 

The physical width of a solitary wave can be defined by $\ell_\s =2|\zeta|$
such that $(v(\zeta)-v_0)/h_\s = e^{-S} \ll 1$,
for some choice of $S>0$. 
We will choose $S=\S$, which captures all of the hump-shaped part of the solitary waves.
From the asymptotic tail \eqref{expdecay} of $v(\zeta)$, 
this yields  
\begin{equation}\label{width}
\ell_\s = 2(\zeta_0 + \S l_\s) 
= \tfrac{2}{\sqrt{3}}R\hat\ell(|\V|/c,v_0) , 
\end{equation}
where
\begin{equation}\label{hatell}
\hat\ell(|\V|/c,v_0) = \sqrt{\tfrac{k(k+1)}{2}}\int_{v_0}^{v_1} \frac{\hat F(v,v_0;|\V|/c)}{v-v_0} \,dv + \S/\sqrt{(\V^2/c^2) v_0^{1-k} - 1}
\end{equation}
is an explicit dimensionless scale depending on $v_0$ and $\V$. 
Note that for $|x-\V t| \gg \ell_\s/2$ the amplitude of the solitary wave 
decays exponentially with respect to the physical length scale \eqref{decayscale}. 

For a fixed wave speed $\V$,
the physical width has the properties 
$\ell_\s\to\infty$ as $v_0\to v^*$,
and 
$\ell_\s\to \sqrt{\tfrac{2}{3}}\tfrac{\sqrt{k(k+1)}}{k-1} \pi R$ as $v_0\to 0$.
The first property is an immediate consequence of
the limit $\S/\sqrt{(v_0/v^*)^{1-k} - 1} \to \infty$ 
for the algebraic term in expression \eqref{hatell},
while the integral term goes to $0$ due to $v_1\to v^*$. 
To derive the second property, 
we note first that the algebraic term in expression \eqref{hatell} goes to $0$,
and the integral term reduces to 
$\int_{0}^{\lambda} (\hat F(v,0;|\V|/c)/v) \,dv$,
where $\hat F(v,0;|\V|/c)= (v/\lambda)^\frac{k-1}{2}/\sqrt{\hat A(v,0;|\V|/c)}$
with $\hat A(v,0;|\V|/c) = 1- (v/\lambda)^{k-1}$. 
We can directly evaluate this integral to get $\tfrac{\pi}{k-1}$,
which gives $\hat\ell \to \sqrt{\tfrac{k(k+1)}{2}}\tfrac{\pi}{k-1}$.

Since the LWHC wave equation \eqref{LWCE-u} is valid only for long wavelength waves,
we must have $\ell_\s\gg 2R$, 
which yields the condition $\tfrac{\sqrt{k(k+1)}}{k-1} \gg \sqrt{6}/\pi$. 
If we impose a lower bound $\ell_\s\gtrsim 4R$, 
then this implies $\tfrac{\sqrt{k(k+1)}}{k-1} \gtrsim 2\sqrt{6}/\pi$,
which holds for all $1<k\lesssim 3.6$.

\subsection{Impulse and energy}

In experiments on pre-compressed chains,
compressive waves are generated by striking an end particle in the chain.
This corresponds to imparting a specified total impulse and total energy 
to the discrete system. 
For the continuum system, described by the LWHC wave equation \eqref{LWCE-u}, 
the total impulse and the total energy of travelling waves $v=v(\zeta)$ 
for the total strain \eqref{strain} in terms of $\zeta = x-\V t$ 
are given by the respective conserved integrals
\begin{equation}\label{unreg_impulse}
\I = \int_{-\infty}^{\infty} \rho\V v\,d\zeta
\end{equation}
and
\begin{equation}\label{unreg_energy}
\E = \int_{-\infty}^{\infty} \rho\big( \tfrac{1}{2}\V^2 v^2 + c^2 ( \tfrac{1}{k(k+1)} v^{k+1} -\tfrac{1}{6}R^2 v^{k-1} {v'}^2 ) \big) \,d\zeta
\end{equation}
which arise directly from the conservation laws \eqref{TX3} and \eqref{TX2}. 
Here $\rho$ is the continuum mass density \eqref{massdens}. 
Both of these integrals are divergent for solitary waves,
because $v(\zeta)$ has a non-zero asymptotic tail $v\to v_0>0$ as $|\zeta|\to \infty$. 
However, we can remove the divergent contribution by subtracting 
the respective impulse density and energy density obtained for $v=v_0$. 
This yields the regularized impulse integral 
\begin{equation}\label{reg_impulse}
\I_\s = \int_{-\infty}^{\infty} \rho\V (v-v_0)\,d\zeta
\end{equation}
and the regularized energy integral 
\begin{equation}\label{reg_energy}
\E_\s = \int_{-\infty}^{\infty} \rho\big( \tfrac{1}{2}\V^2 (v^2-v_0^2) + c^2 ( \tfrac{1}{k(k+1)} (v^{k+1}-v_0^{k+1}) -\tfrac{1}{6}R^2 v^{k-1} {v'}^2 ) \big) \,d\zeta ,
\end{equation}
both of which will be finite for all solitary waves. 
These regularized integrals can be simplified 
by using the method explained in \Ref{PrzAnc}.
We obtain 
\begin{equation}\label{impulse}
\I_\s = \tfrac{2}{\sqrt{3}} \rho Rc\, \sgn(\V)\hat\I(|\V|/c,v_0)
\end{equation}
and 
\begin{equation}\label{energy}
\E_\s = \tfrac{1}{\sqrt{3}} \rho Rc|\V|\hat\E(|\V|/c,v_0)
\end{equation}
where 
\begin{align}
\hat\I(|\V|/c,v_0) & = \int_{v_0}^{v_1} \frac{\sqrt{v^{k-1}}}{\sqrt{\hat A(v,v_0;|\V|/c)}}\,dv
\label{impulse_integral}
\\
\hat\E(|\V|/c,v_0) & = \int_{v_0}^{v_1} \sqrt{v^{k-1}}\left( \frac{v^2-v_0^2 + \tfrac{2}{k(k+1)}(c/\V)^2( v^{k+1}-v_0^{k+1} )}{(v-v_0)\sqrt{\hat A(v,v_0;|\V|/c)}} - (v-v_0)\sqrt{\hat A(v,v_0;|\V|/c)} \right)\,dv
%\int_{v_0}^{v_1} v^{(k-1)/2}\left( \frac{v+v_0 + \tfrac{2}{k(k+1)}(c^2/\V^2)( \frac{v^{k+1}-v_0^{k+1}}{v-v_0} )}{\sqrt{\hat A(v,v_0;\V)}} - (v-v_0)\sqrt{\hat A(v,v_0;\V)} \right)\,dv
\label{energy_integral}
\end{align}
are explicit dimensionless integrals depending on $v_0$ and $|\V|/c$.

\subsection{Speed dependence of height, width, impulse, energy}

When the background strain $v_0>0$ is fixed,
all solitary waves comprise a one-parameter family 
given in terms of the wave speed $\V$ in the range \eqref{minspeed}. 
Note that fixing $v_0$ corresponds to fixing the sound speed \eqref{soundspeed} 
(where $k$ is taken to be fixed). 
Then the impulse, energy, height and width of a solitary wave 
will depend solely on the ratio $|\V|/c_0$. 

\begin{figure}[!h]
\centering
\includegraphics[width=0.45\textwidth]{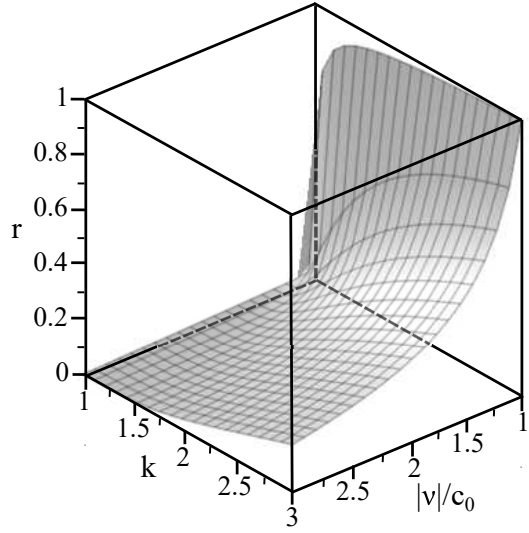}
\caption{
Strain ratio $r$, defined by \Eqref{rroot}, as a function of the speed ratio $|\V|/c_0$ and the exponent $k$ in the Hertz potential. 
}
\label{fig:strainratio_plots}
\end{figure}

To understand the specific dependence of these physical properties on $|\V|/c_0$, 
we first need to determine the peak strain $v_1$ in terms of $c_0$ and $\V$. 
The algebraic equation \eqref{v1Aeqn} for $v_1$ can be expressed in the explicit form 
\begin{equation}\label{v1root}
0= \tfrac{k(k+1)}{2}(\V^2/c^2) (v_1-v_0)^2 -v_1^{k+1} +(k+1)v_0^{k}v_1 -kv_0^{k+1} .
\end{equation} 
For convenience, we will write 
\begin{equation}\label{strainratio}
r=v_0/v_1 , 
\end{equation} 
which has the range
\begin{equation}\label{rrange}
0 <r < 1
\end{equation} 
given by the intervals \eqref{v0v1interval}. 
In terms of this strain ratio \eqref{strainratio}, 
the algebraic equation \eqref{v1root} is given by 
\begin{equation}\label{rroot}
0= k(r-1)r^{k-1}\big( \tfrac{k+1}{2}(\V^2/c_0^2) (r-1) -r \big) + r^{k} -1 
\end{equation} 
by use of relation \eqref{v0c0rel}.
Hence, for a given value of $k$, 
$r$ is a function only of the speed ratio $|\V|/c_0$. 
This function is shown in \figref{fig:strainratio_plots}. 

It is useful to observe that scaling $c \to s^\frac{k-1}{2}c$ (with $s\neq 0$)
implies $v_0\to s v_0$ and $v_1\to s v_1$,
and thus
\begin{align}
& \hat F(s v,s v_0;|\V|/c) =s^\frac{k-1}{2}\hat F(v,v_0;s^\frac{1-k}{2}|\V|/c), 
\\
& \hat\I(|\V|/c,sv_0) = s^\frac{k+1}{2}\hat\I(s^\frac{1-k}{2}|\V|/c,v_0) , 
\\
& \hat\E(|\V|/c,sv_0) = s^\frac{k+3}{2}\hat\E(s^\frac{1-k}{2}|\V|/c,v_0) . 
\end{align}
Then, using these scaling properties combined with relation \eqref{v0c0rel}, 
we have the following properties of solitary waves
expressed in terms of the strain ratio \eqref{strainratio}. 

The height \eqref{height} is given by 
\begin{equation}\label{height:r}
h_\s = (c_0/c)^{\frac{2}{k-1}}(1/r-1) ,
\end{equation}
while the width \eqref{width}, impulse \eqref{impulse} and energy \eqref{energy}
are given in terms of the integrals 
\begin{align}
\hat\ell &= \sqrt{\tfrac{k(k+1)}{2}} \int_{1}^{1/r} \frac{\hat F(z,1;|\V|/c_0)}{z-1} \,dz +\S/\sqrt{\V^2/c_0^2 -1} , 
\label{width_integral:r}
\\
\hat\I & = (\tfrac{c_0}{c})^\frac{k+1}{k-1} \int_{1}^{1/r} \frac{\sqrt{z^{k-1}}}{\sqrt{\hat A(z,1;|\V|/c_0)}}\,dz , 
\label{impulse_integral:r}
\\
\hat\E & = (\tfrac{c_0}{c})^\frac{k+3}{k-1}  \int_{1}^{1/r} \sqrt{z^{k-1}}\left( \frac{z^2-1 + \tfrac{2}{k(k+1)}(c_0/\V)^2( z^{k+1}-1 )}{(z-1)\sqrt{\hat A(z,1;|\V|/c_0)}} - (z-1)\sqrt{\hat A(z,1;|\V|/c_0)} \right)\,dz . 
\label{energy_integral:r}
\end{align}
Here 
\begin{equation}\label{hatF}
\sqrt{\tfrac{k(k+1)}{2}} \hat F(z,1;|\V|/c_0) = (c_0/|\V|) \Bigg( \frac{\sqrt{z^{k-1}}}{\sqrt{\hat A(z,1;|\V|/c_0)}} - \frac{1}{\sqrt{\hat A(1,1;|\V|/c_0)}} \Bigg) , 
\end{equation}
and 
\begin{equation}\label{hatA}
\begin{aligned}
\hat A(z,1;|\V|/c_0) & = 1- \tfrac{2}{k(k+1)}(c_0/\V)^2 (k-z\partial_z)\Big(\frac{z^k-1}{z-1}\Big) ,
\\
\hat A(1,1;|\V|/c_0) & = 1- (c_0/\V)^2 .
\end{aligned}
\end{equation}

\begin{figure}[!h]
\centering
\includegraphics[width=0.95\textwidth]{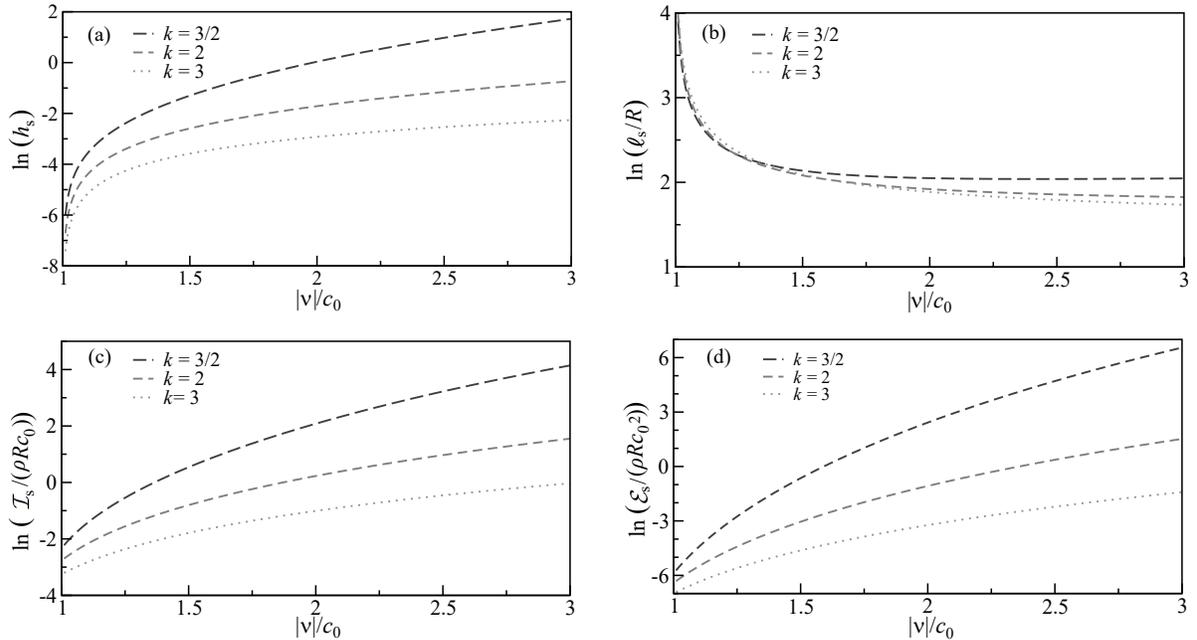}
\caption{
Properties of the solitary wave solutions presented in Section~\ref{sec:solitary-examples}, using a fixed background strain $v_0 = 0.02$. 
(a) Solitary wave height, \Eqref{height:r};
(b) solitary wave width, \Eqref{width_integral:r};
(c) impulse carried by solitary waves, \Eqref{impulse_integral:r};
(d) energy carried by solitary waves, \Eqref{energy_integral:r}.
}
\label{fig:height_width_impulse_plots}
\end{figure}

The height, width, impulse, and energy are shown in \figref{fig:height_width_impulse_plots}
for the explicit solitary wave solutions presented in Section~\ref{sec:solitary-examples}, where
the background strain $v_0$ has been fixed to be a typical value 
considered in experiments. 
It is evident from \figref{fig:height_width_impulse_plots}(a) that 
the physical height $h_\s$ of a solitary wave 
increases with increasing $|\V|/c_0$ (i.e., with decreasing $r$). 
Moreover, for a given speed ratio $|\V|/c_0$, 
$h_\s$ decreases with increasing $k$. 
In contrast, for a given value of $k$, 
the width $\ell_\s$ of a solitary wave increases with decreasing $|\V|/c_0$, 
implying that the width becomes larger when the initial pre-compression is increased. 

From \figref{fig:height_width_impulse_plots}(b), 
we see that, when $|\V|/c_0 \gtrsim 1.5$, 
the width $\ell_\s$ decreases with increasing $k$ for a given $|\V|/c_0$.  
In contrast, when $|\V|/c_0 \lesssim 1.5$,
this trend is inverted, with $\ell_\s$ increasing as $k$ increases. 
This means that solitary waves are wider for a larger values of $k$ 
when the wave speed $|\V|$ is less than $1.5$ times the sound speed $c_0$.  

Similarly to the properties of the height and width, 
both the impulse and energy, \figref{fig:height_width_impulse_plots}(c) and (d), 
increase with increasing $|\V|/c_0$ for a given value of $k$. 
When $k$ and $|\V|/c_0$ are fixed, 
the impulse, and energy each increase if the background strain $v_0$ is increased,
which can be inferred from expressions \eqref{impulse_integral:r} and \eqref{energy_integral:r}. 
Physically, this means that a larger impulse and energy is needed to obtain waves with the same speed ratio when the initial pre-compression is increased. 

We will now consider in more detail two interesting cases:
wave speeds $|\V|$ close to the sound speed \eqref{soundspeed}; 
wave speeds $|\V|$ much larger than the sound speed \eqref{soundspeed}. 
Note, since the background strain $v_0$ is fixed, 
specifying the speed ratio $|\V|/c_0$ corresponds to specifying $v^*$ through 
the relation $|\V|/c_0 = (v^*/v_0)^{(k-1)/2}$.

\subsection{Slightly supersonic solitary waves}

When $|\V|$ is close to $c_0$, 
a solitary wave is slightly supersonic. 
To examine its properties, 
we first observe that $v^*$ will be close to $v_0$, 
whereby the strain ratio $r$ will be close to $1$
as seen from \figref{fig:strainratio_plots}. 
By expanding equation \eqref{rroot} for $r$ in a series in $|\V|/c_0 - 1$ around $r=1$, 
we obtain the asymptotic expression 
\begin{equation}\label{r_near1}
r\simeq 1- \tfrac{6}{k-1}(|\V|/c_0 - 1) 
\end{equation}
valid for $|\V|/c_0 -1 \ll 1$. 

A similar asymptotic expansion of the height expression \eqref{height:r}
and the width integral \eqref{width_integral:r}
yields 
$h_\s  \simeq \tfrac{6}{k-1} (c_0/c)^\frac{2}{k-1} (|\V|/c_0 - 1)$
and 
$\hat\ell \simeq \tfrac{\S}{\sqrt{2}}\big/\sqrt{|\V|/c_0 - 1}$. 
(see Appendix~\ref{appendix:expansion} for details). 
Thus, we see that the width \eqref{width} of a slightly supersonic solitary wave 
becomes large in comparison to the particle size $2R$, 
\begin{equation}\label{width:r_near_1}
\ell_\s/(2R) \simeq \tfrac{\S}{\sqrt{6}} \big/\sqrt{|\V|/c_0 - 1} \gg 1, 
\end{equation}
while the height becomes small compared to the background strain $v_0$, 
\begin{equation}\label{height:r_near_1}
h_\s/v_0  \simeq \tfrac{6}{k-1} (|\V|/c_0 - 1) \ll \tfrac{6}{k-1} . 
\end{equation}
In particular, the height and width satisfy the proportionality relationship
\begin{equation}\label{heightwidthrel:r_near_1}
h_\s \propto 1/\ell_\s^2 .
\end{equation}

The impulse integral \eqref{impulse_integral:r} 
and the energy integral \eqref{energy_integral:r}
have the respective asymptotic expansions 
$\hat\I \simeq \tfrac{3\sqrt{2}}{k-1}(\tfrac{c_0}{c})^\frac{k+1}{k-1}\sqrt{|\V|/c_0 - 1}$
and 
$\hat\E \simeq \tfrac{6\sqrt{2}(k+1)}{k(k-1)}(\tfrac{c_0}{c})^\frac{k+3}{k-1}\sqrt{|\V|/c_0 - 1}$
(see Appendix~\ref{appendix:expansion} for details). 
Thus the impulse \eqref{impulse} and the energy \eqref{energy} of a slightly supersonic solitary wave 
are given by 
\begin{equation}\label{impulse:r_near_1}
\I_\s \simeq \tfrac{2\sqrt{6}}{k-1} \rho Rc_0 (c_0/c)^\frac{2}{k-1} \sgn(\V)\sqrt{|\V|/c_0 - 1} ,
\end{equation}
and 
\begin{equation}\label{energy:r_near_1}
\E_\s \simeq \tfrac{2\sqrt{6}(k+1)}{k(k-1)} \rho Rc_0^2(c_0/c)^\frac{4}{k-1} \sqrt{|\V|/c_0 - 1} . 
\end{equation}
In particular, these two quantities are related by 
$\E_\s/\I_\s \simeq (1+\tfrac{1}{k})(c_0/c)^\frac{2}{k-1}c_0\,\sgn(\V)$,
which depends only on the sound speed $c_0$ and the direction of the solitary wave
(in addition to the constant $c$). 

As a limiting case, 
for $|\V|\to c_0$, 
we have $\ell_\s/(2R)\to \infty$, 
$h_\s/v_0 \to 0$, $\I_\s \to 0$ and $\E_\s \to 0$.

\subsection{Highly supersonic solitary waves}

When $|\V|$ is much larger than $c_0$, 
a solitary wave is highly supersonic. 
This implies $v^*$ will be much greater than $v_0$, 
whereby the strain ratio $r$ will be much less than $1$
as seen from \figref{fig:strainratio_plots}. 
Consequently, 
by expanding equation \eqref{rroot} for $r$ in a series in $(c_0/|\V|)^\frac{2}{k-1}$ around $r=0$, 
we obtain the asymptotic expression 
\begin{equation}\label{r_near0}
r\simeq (\tfrac{2}{k(k+1)}c_0^2/\V^2)^\frac{1}{k-1}
\end{equation}
valid for $|\V|/c_0 \gg 1$. 

The properties of highly supersonic solitary waves can be determined using 
this asymptotic form of the strain ratio. 

First, we asymptotically expand the height \eqref{height:r},
which yields 
\begin{equation}\label{height:strong}
h_\s \simeq (\tfrac{k(k+1)}{2})^\frac{1}{k-1} (|\V|/c)^\frac{2}{k-1} . 
\end{equation}
Hence, compared to the background strain $v_0$, 
the height becomes large, 
\begin{equation}\label{height:r_gtr_1}
h_\s/v_0 \simeq (\tfrac{k(k+1)}{2})^\frac{1}{k-1} (|\V|/c_0)^\frac{2}{k-1} \gg (\tfrac{k(k+1)}{2})^\frac{1}{k-1} .
\end{equation}

Next, an asymptotic expansion of the width integral \eqref{width_integral:r} 
yields 
\begin{equation}\label{hatell:r_gtr_1}
\hat\ell \simeq \tfrac{\sqrt{k(k+1)}}{\sqrt{2}(k-1)}\pi
\end{equation}
(as shown in Appendix~\ref{appendix:expansion}). 
Thus, the asymptotic expression for the width \eqref{width} is given by 
$\ell_\s \simeq \sqrt{\tfrac{2}{3}}\tfrac{\sqrt{k(k+1)}}{k-1}\pi R$.
In comparison to the particle size $2R$, 
the width is a finite multiple 
\begin{equation}\label{width:r_gtr_1}
\ell_\s/(2R) \simeq \tfrac{1}{\sqrt{6}}\tfrac{\sqrt{k(k+1)}}{k-1}\pi , 
\end{equation}
which depends only on $k$. 
This multiple is a decreasing function of $k$, such that 
$\ell_\s/(2R)\to  \tfrac{1}{\sqrt{6}}\pi $ ($\simeq 1.28$) as $k\to\infty$
and $\ell_\s/(2R)\to  \infty$ as $k\to 1$. 
The trend for large $k$ is in agreement with previous experimental findings \cite{Sen2008} for discrete systems, where it has been reported that $\ell_\s/(2R)$ tended to $1$. 

In the physically important case $k=\tfrac{3}{2}$, 
the width is $\ell_\s/(2R)\simeq \tfrac{\sqrt{5}}{\sqrt{2}}\pi$ ($\simeq 4.97$). 
This agrees with the experimentally measured value \cite{Daraio2006Tunability} of approximately $10R$ 
for the physical width of solitary waves in weakly compressed granular chains 
when the contact geometry of the particles is elliptical, corresponding to $k=\tfrac{3}{2}$. 

Similarly, 
we find that the impulse integral \eqref{impulse_integral:r} 
and the energy integral \eqref{energy_integral:r}
have the respective asymptotic expansions
$\hat I \simeq \lambda^\frac{k+1}{2} \sqrt{\pi} \Gamma(\tfrac{1}{2} +\tfrac{1}{k-1})/\Gamma(\tfrac{1}{k-1})$
and 
$\hat E \simeq \lambda^\frac{k+3}{2}\tfrac{1}{4}\sqrt{\pi} \tfrac{k+3}{k+1}\Gamma(\tfrac{1}{2} + \tfrac{2}{k-1})/\Gamma(\tfrac{2}{k-1})$
(derived in Appendix~\ref{appendix:expansion}). 
Thus the impulse \eqref{impulse} and the energy \eqref{energy} of a highly supersonic solitary wave 
are given by 
\begin{equation}\label{impulse:r_gtr_1}
\I_\s \simeq \tfrac{2}{\sqrt{3}}\sqrt{\pi} K_1 \rho \V  (|\V|/c)^\frac{2}{k-1} R ,
\quad
K_1 = \sqrt{\tfrac{k(k+1)}{2}}^\frac{k+1}{k-1} \tfrac{\Gamma(\frac{1}{2}+\frac{1}{k-1})}{\Gamma(\frac{1}{k-1})} , 
\end{equation}
and 
\begin{equation}\label{energy:r_gtr_1}
\E_\s \simeq \tfrac{1}{4\sqrt{3}}\sqrt{\pi} K_2 \rho \V^2  (|\V|/c)^\frac{4}{k-1} R ,
\quad
K_2 = \tfrac{k+3}{k+1}\sqrt{\tfrac{k(k+1)}{2}}^\frac{k+3}{k-1} \tfrac{\Gamma(\frac{1}{2} + \frac{2}{k-1})}{\Gamma(\frac{2}{k-1})} .
\end{equation}
In particular, these two quantities are related by 
$\E_\s/\I_\s \simeq \tfrac{1}{8}(1+\tfrac{2}{k+1}) \hat K \V h_\s$ in terms of the height and the speed,
with $\hat K = \Gamma(\tfrac{1}{k-1})\Gamma(\tfrac{1}{2}+\tfrac{2}{k-1})/\big(\Gamma(\tfrac{2}{k-1})\Gamma(\tfrac{1}{2}+\tfrac{1}{k-1})\big)$.

\section{Comparison of solitary waves across different nonlinear regimes}\label{compare}

In discrete systems of pre-compressed particles, 
the features of compressive and rarefactive pulses 
are very dependent on the size of the dynamical overlap of adjacent particles 
compared to the size of the pre-compression. 
This dependence is measured in the continuum limit 
by the ratio of the dynamical strain \eqref{dynstrain} to the background strain \eqref{initialstrain}. 
Specifically, the dynamics of compressive long-wavelength pulses (in the continuum limit) is 
weakly nonlinear when the strain ratio satisfies the condition \eqref{weaknonlin:V},
or strongly nonlinear when the strain ratio satisfies the condition \eqref{strongnonlin:V}. 

We will now explore how the main features of solitary waves 
described by the LWHC wave equation \eqref{LWCE-u} 
differs across these two different nonlinearity regimes. 

To begin, 
we note that the peak total strain $v_1$ in a solitary wave \eqref{phys_solitarywave} 
is a function of the background strain $v_0$ and the wave speed $\V$,
as given by equation \eqref{v1Aeqn},
where $v_0>0$ and $\V >c_0$. 
Since the degree of dynamical nonlinearity is essentially determined by the ratio $1/r=v_1/v_0$, 
which has the range \eqref{rrange}, 
we can regard solitary waves as being determined by $r$ and $\V$ 
as two independent parameters. 
Specifically, from the algebraic equation \eqref{rroot} for $r$, 
combined with relation \eqref{v0c0rel} for $v_0$, 
we have
\begin{equation}\label{v0v1_r}
v_0 = \Big( \tfrac{k(k+1)}{2} \frac{(\V^2/c^2) (1-r)^2}{r^{1-k} -k(1-r)r-r} \Big)^\frac{1}{k-1},
\quad
v_1 = v_0/r ,
\end{equation} 
which explicitly expresses the background strain and the peak strain as functions of $r$ and $\V$. 
Since $v_0$ determines the sound speed $c_0$ through relation \eqref{v0c0rel},
note that expression \eqref{v0v1_r} yields
\begin{equation}\label{c0_r}
c_0 =  \sqrt{\tfrac{k(k+1)}{2}} \frac{|\V| (1-r)}{\sqrt{r^{1-k} -k(1-r)r-r}} 
\end{equation}
as a function of $r$ and $|\V|$. 
Consequently, weak nonlinearity arises when solitary waves are slightly supersonic, 
whereas strong nonlinearity occurs when solitary waves are highly supersonic,
which we will show explicitly from the sound speed equation \eqref{c0_r} later. 

There are two ways in which we can compare different solitary waves:
one way is to look at all solitary waves that have the same physical speed, 
namely a family of waves parameterized by $r$ with $\V$ being fixed;
another way is to consider all solitary waves that have the same impulse $\I_\s$, 
since fixing the impulse will determine $\V$ in terms of $v_0$, 
which gives a family of waves parameterized by $r$ with $\I_\s$ being fixed. 

For comparisons of solitary waves,
a useful quantity to study is the scaled wave profile $(v-v_0)/h_\s$ 
as a function of the scaled travelling wave variable $\zeta/\ell_\s = (x-\V t)/\ell_\s$,
where $h_\s$ is the peak height of the wave and $\ell_\s$ is the width of the wave. 
The scaled profile has a range $0$ to $1$, 
while the scaled travelling wave variable can be taken in the interval $(-\tfrac{1}{2},\tfrac{1}{2})$ 
to exclude the asymptotic tail of a solitary wave.

\subsection{Strongly nonlinear regime}

In the strongly nonlinear regime \eqref{strongnonlin:V},
the peak total strain $v_1$ in a solitary wave satisfies $v_1 \gg v_0$.
In terms of the strain ratio \eqref{strainratio}, 
this regime is characterized by $1/r \gg 1$, 
namely $r=v_0/v_1$ is small. 
Using the sound speed equation \eqref{c0_r}, we then have 
$c_0/|\V| \simeq \sqrt{\tfrac{k(k+1)}{2}} r^\frac{k-1}{2}$. 
Hence, this regime coincides with the case of solitary waves that are highly supersonic 
\begin{equation}\label{speedratio:strong}
|\V| \simeq \sqrt{\tfrac{2}{k(k+1)}} r^\frac{1-k}{2} c_0 \gg c_0
\end{equation}
where $r\ll 1$. 
It will be useful to note 
\begin{equation}\label{v1:strong}
v_1\simeq (\tfrac{k(k+1)}{2})^\frac{1}{k-1}(|\V|/c)^\frac{2}{k-1}
\end{equation}
holds in this regime, 
as shown by combining the relations \eqref{speedratio:strong} and \eqref{v0c0rel}.

The main properties of strongly nonlinear solitary waves in terms of $r$ and $\V$ 
are given by the expressions \eqref{height:strong}, \eqref{width:r_gtr_1}, \eqref{impulse:r_gtr_1}, \eqref{energy:r_gtr_1}
for the height, width, impulse, and energy in the highly supersonic case. 
To leading order in $r$, 
these properties are asymptotically independent of $r$. 
In particular, the width \eqref{width:r_gtr_1} is a constant 
(which depends on $k$), 
and thus strongly nonlinear solitary waves of different speeds have approximately the same width. 
In contrast, 
the height \eqref{height:strong}, impulse \eqref{impulse:r_gtr_1}, and energy \eqref{energy:r_gtr_1} are functions of $\V$. 

The expressions for the height \eqref{height:strong} and width \eqref{width:r_gtr_1} 
of strongly nonlinear solitary waves 
are the same as the height and width of the compact solution \eqref{compactonsol}
\begin{equation}\label{phys_compactonsol}
v_\c = (\tfrac{k(k+1)}{2})^\frac{1}{k-1} (|\V|/c)^\frac{2}{k-1} \cos\big( (x-\V t)\pi/l_\c \big)^\frac{2}{k-1} \Theta\big(\pi(\tfrac{1}{2}-|x-\V t|/l_\c)\big)
\end{equation}
which exists in the limiting case $r=0$. 
Here $\Theta(z)$ denotes the Heaviside step function,
and 
\begin{equation}\label{coswidthscale}
l_\c = \tfrac{\sqrt{2k(k+1)}}{\sqrt{3}(k-1)} \pi R 
\end{equation}
is a constant (which depends on  $k$ and $R$). 
The presence of the step function causes $v$ to vanish for 
$|x-\V t|\geq \tfrac{1}{2}l_\c$,
whereby $v$ is compactly supported as a function of the travelling wave variable $x-\V t$.

At any fixed value of $r$, 
the width \eqref{width:r_gtr_1} and height \eqref{height:strong}
decrease with increasing $k$. 
Specifically, 
the width $\ell_\s\simeq \ell_\c$ approaches a limiting size $\tfrac{2}{\sqrt{6}}\pi R $ ($\simeq 2.56R$) 
which corresponds to a multiple $1.28$ of a particle diameter,
while the height $h_\s$ approaches the limit $1$. 

As discussed in Section~\ref{sec:solitary-examples}, 
the compacton \eqref{phys_compactonsol} 
is an actual solution of the LWHC wave equation \eqref{LWCE-u} 
only for $1<k<\tfrac{5}{3}$,
since the cutoff imposed by the step function lacks sufficient differentiability 
when $k\geq \tfrac{5}{3}$.
Moreover, this solution holds only in the case when the background strain \eqref{initialstrain} is zero,
which describes a continuum system with \emph{no} pre-compression, 
$v_0=0$. 

In contrast, strongly nonlinear solitary waves have non-zero tails for all $0<r\ll 1$. 
We will now show that the shape of strongly nonlinear solitary waves 
near their peak is approximated by the upper part of the compacton profile:
\begin{equation}\label{archapprxsoln}
v/v_0 \simeq (\tfrac{k(k+1)}{2})^\frac{1}{k-1} (|\V|/c_0)^\frac{2}{k-1} \cos\big( (x-\V t)\pi/l_\c \big)^\frac{2}{k-1}
\end{equation}
for 
\begin{equation} 
|x-\V t| \ll \tfrac{1}{2}l_\c  ,
\end{equation}
where $l_\c\simeq l_\s$. 
In particular, the scaled wave profile as a function of the scaled travelling wave variable $|\zeta|/\ell_\s$ is approximately given by 
$(v-v_0)/h_\s \simeq \cos(|\zeta|\pi/\ell_\s)^\frac{2}{k-1}$
in the interval $0\leq |\zeta|/\ell_\s \ll \tfrac{1}{2}$,
where the cosine function is not close to $0$. 

\begin{figure}[!h]
\centering
\includegraphics[width=\textwidth]{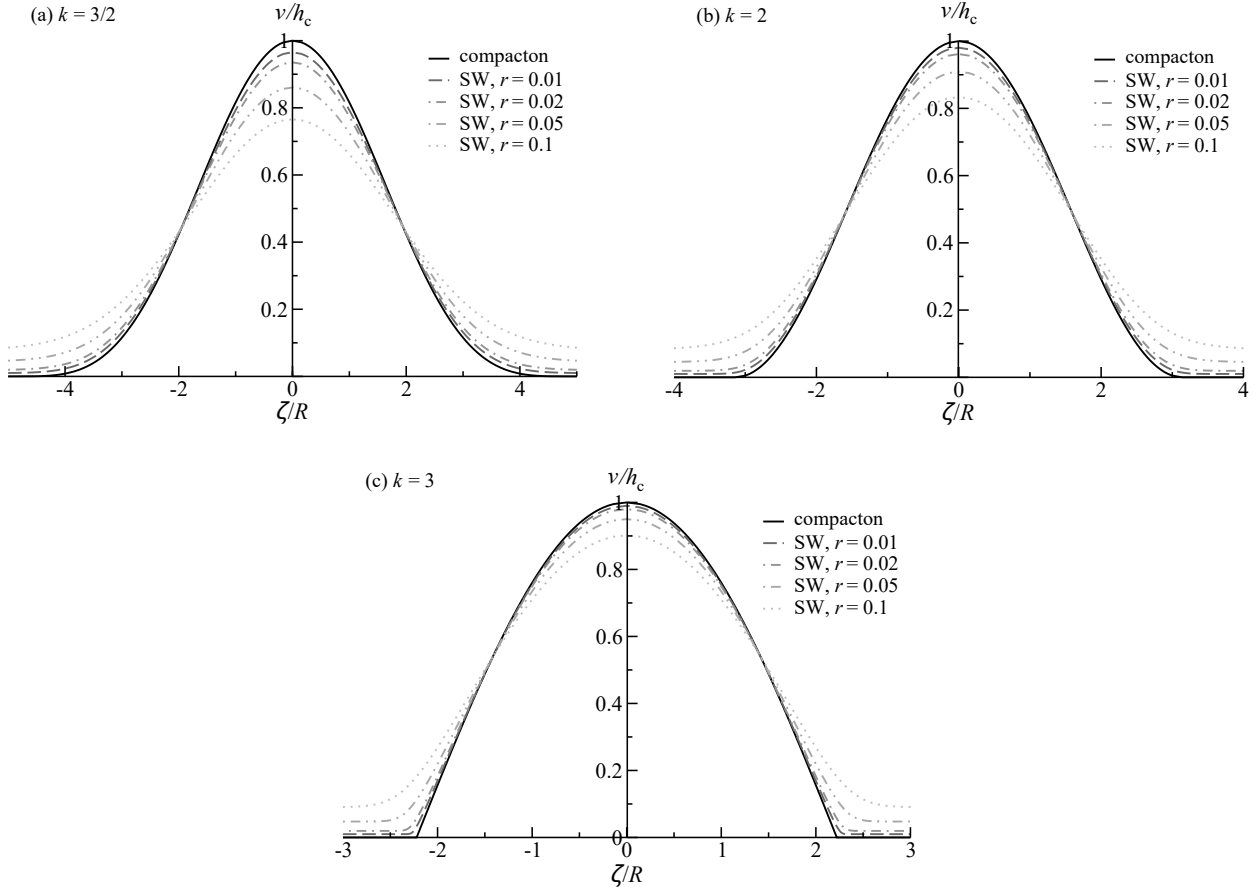}
\caption{
Comparison between the compacton solution \eqref{phys_compactonsol} and the solitary wave (SW) solutions given by the quadrature \eqref{phys_solitarywave} for various values of $r \ll 1$. 
(a) $k=3/2$; 
(b) $k=2$; 
(c) $k=3$.  
Here, $h_\c$ is the compacton height, given by \Eqref{height:strong} (see \Eqref{phys_compactonsol}), which is equivalent to \Eqref{scaling}. 
}
\label{fig:compacton_comparison_plots}
\end{figure}

This approximation \eqref{archapprxsoln} holds only for the part of the solitary wave 
where the amplitude is $v/v_0 \gg 1$. 
When instead the amplitude is $v/v_0 \simeq 1$, 
the tail of the solitary wave is given by the decaying exponential \eqref{expdecay}
with $\zeta_0\simeq \tfrac{1}{\sqrt{3}}R\hat\ell$. 
For the part of the wave profile where $v/v_0$ is neither large nor close to $1$,
no explicit approximation appears to be possible because of the strong nonlinearity. 
A comparison of the exact solitary wave solution with $r\ll 1$ 
and the compacton solution for $r=0$ is shown in \figref{fig:compacton_comparison_plots}
for $k=3/2, 2, 3$.

It is clear from \figref{fig:compacton_comparison_plots} 
that the compacton well-approximates a solitary wave when $r \ll 1$, 
and that the approximation becomes worse with increasing $r$.  
For a given value of $r$, 
the approximation is better for larger values of $k$. 
While the compacton is a good approximation for $k=2$ and $k=3$ solitary waves when $r \ll 1$, 
it should be noted that the compacton is not an actual solution to the LWHC when $k>5/3$,
as explained in Section~\ref{sec:compacton}. 
Contact geometries with $k>5/3$ occur in several physically interesting discrete systems,
particularly when the macroscopic particles in the system have rough contact surfaces 
\cite{Spence1968,Goddard1990,Persson2006}. 

Interestingly, for any $k>1$, 
we can use the exact quadrature formula \eqref{phys_solitarywave} for solitary waves
to derive an approximate quadrature that describes the entire solitary wave profile 
and that can be evaluated explicitly when $v/v_0 \gg 1$ to obtain the approximation \eqref{archapprxsoln} near the peak,
as well as when $v/v_0\simeq 1$ to obtain the approximation \eqref{expdecay} for the tail. 

We start by writing out the full form of the quadrature \eqref{phys_solitarywave}, 
specifically the square-root term in the integrand. 
This term is given by 
$\hat A(v,v_0;|\V|/c)= 1- \lambda^{1-k}\partial_{v_0}\big(v_0(v^k-v_0^k)/(v-v_0)\big)$,
which has the approximation 
$\hat A(v,v_0;|\V|/c) \simeq 1- \lambda^{1-k}v^{k-1}$ 
to leading-order in $v_0/v$,
with $\lambda$ given by the relation \eqref{vstar}. 
Hence the quadrature can be approximated by 
 \begin{equation}\label{strong:apprxsol}
\int_{v}^{v_1} \frac{\sqrt{v^{k-1}}\,dv}{(v-v_0)\sqrt{1-\lambda v^{k-1}}}
\simeq \sqrt{3}(|\V|/c) |\zeta|/R ,
\end{equation}
where we keep the exact pole term $v-v_0$ in the integrand 
because it yields the tail of the solitary wave to leading-order in $v_0/v$. 
This quadrature \eqref{strong:apprxsol} provides a useful approximation that encompasses the entire wave profile. 
It reduces to the exact compacton solution if we put $v_0=0$. 
Moreover, for $v_0>0$, if we restrict $v$ so that $v/v_0 \gg 1$,
which holds near the peak of the wave profile, 
then we obtain the approximation \eqref{archapprxsoln}
given by the upper part of the arch of the compacton solution. 

Unfortunately, the approximate quadrature \eqref{strong:apprxsol} 
cannot be evaluated explicitly for arbitrary $k>1$ when $v_0>0$. 
But we can evaluate it for the same special values of $k$ 
for which the exact quadrature was evaluated in Section~\ref{sec:solitary-examples}. 

Specifically, when $k=2$, we have 
\begin{equation}\label{strong:apprxsol:k=2}
\arccos(2v/\lambda-1) + \sqrt{v_0/\lambda} \ln\Big(\frac{v_0 + v +2\sqrt{v_0}\sqrt{v(1-v/\lambda)}}{v-v_0}\Big)
\simeq \tfrac{2\pi}{k-1}|\zeta|/l_\c , 
\end{equation}
where we have used relation \eqref{v1:strong} for $v_1$, 
and where we have dropped all terms that are small due to $c_0/|\V| \ll 1$,
which holds in the strongly nonlinear regime. 
When $k=3$, we obtain 
\begin{equation}\label{strong:apprxsol:k=3}
\tfrac{1}{2}\arccos(2v^2/\lambda^2-1) + (v_0/\lambda) \ln\Big(\frac{1-v_0v/\lambda^2 +\sqrt{1-v^2/\lambda^2}}{v-v_0}\Big)
\simeq \tfrac{2\pi}{k-1}|\zeta|/l_\c  . 
\end{equation}
Likewise, when $k=3/2$, we have
\begin{equation}\label{strong:apprxsol:k=3/2}
\begin{aligned}
2\arccos(2\sqrt{v/\lambda}-1) - \sqrt[4]{v_0/\lambda}\bigg( 
& \ln\bigg(\frac{\sqrt{v_0} +\sqrt{v}+2\sqrt[4]{v_0v}\sqrt{1-\sqrt{v/\lambda}}}{v-v_0}\bigg)
\\& 
-\arctan\bigg(\frac{2\sqrt[4]{v_0v}\sqrt{1-\sqrt{v/\lambda}}}{v-v_0}\bigg)  \bigg) 
\simeq \tfrac{2\pi}{k-1}|\zeta|/l_\c  . 
\end{aligned}
\end{equation}

\begin{center}
\begin{figure}[!h]
\centering
\includegraphics[width=1.08\textwidth]{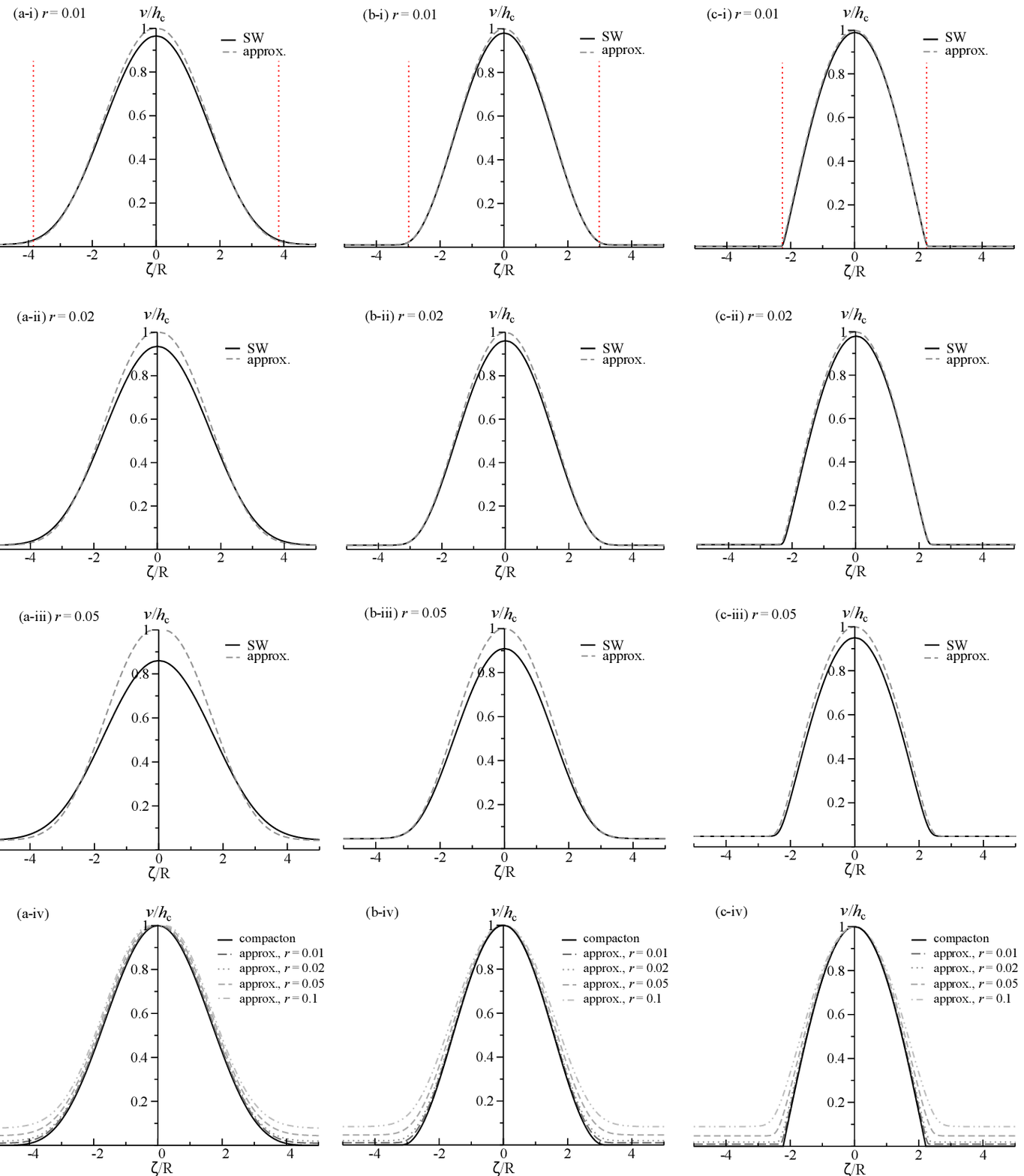}
\caption{
(Color online) Comparison of solutions of the approximate (approx.) quadrature \eqref{strong:apprxsol} with solitary wave (SW) solutions of \eqref{phys_solitarywave}  
for various values of $r\ll 1$.
Column (a) $k=3/2$, \Eqref{strong:apprxsol:k=3/2}; 
column (b) $k=2$, \Eqref{strong:apprxsol:k=2}; 
column (c) $k=3$, \Eqref{strong:apprxsol:k=3}.  
The solitary wave width is indicated by dashed vertical lines in the row (i).
The compacton solution \eqref{phys_compactonsol} is shown for comparison in row (iv). Here, $h_{c}$  is the compacton height, given by \Eqref{height:strong}. 
}
\label{fig:strongly_nonlinear_approx_plots}
\end{figure}
\end{center}

These explicit approximations \eqref{strong:apprxsol:k=2}--\eqref{strong:apprxsol:k=3/2}  
along with the corresponding exact solitary wave solutions given by \eqref{phys_solitarywave} 
for $r \ll 1$
are shown in Fig.~\ref{fig:strongly_nonlinear_approx_plots}. 
The plots indicate a good agreement between
the exact solitary wave solutions and the approximate solitary wave solutions 
when $r$ is very small, 
while the approximation gradually worsens with increasing $r$, 
particularly near the peak of the wave. 
Similar to the compacton solution~\eqref{phys_compactonsol}, 
at fixed $r$ the approximate solutions \eqref{strong:apprxsol:k=2}--\eqref{strong:apprxsol:k=3/2} 
agree better with the exact solitary wave solutions for larger values of $k$. 
The compacton solution is shown for comparison in row (iv) of Fig.~\ref{fig:strongly_nonlinear_approx_plots}.
As illustrated by the plots, for all values of $r$, the shape of the peak of the approximate 
solution is nearly identical to the compacton; 
however, the tail of the approximate solution 
agrees much better with the exact solitary wave solution. 

\subsection{Weakly nonlinear regime}

In the weakly nonlinear regime \eqref{weaknonlin:V},
the peak total strain $v_1$ in a solitary wave satisfies $v_1-v_0\ll v_0$.
This regime is characterized in terms of the strain ratio \eqref{strainratio} by 
$1/r - 1 \ll 1$, 
namely $r=v_0/v_1$ is close to $1$. 
From the sound speed equation \eqref{c0_r}, we then see 
$c_0/|\V| \simeq 1 - \tfrac{k-1}{6}(1-r)$, 
showing that this regime coincides with the case of solitary waves that are slightly supersonic 
\begin{equation}\label{speedratio:weak}
|\V| \simeq (1 + \tfrac{k-1}{6}(1-r))c_0 \gtrsim c_0
\end{equation}
where $1-r\ll 1$. 

We can now obtain the main properties of weakly nonlinear solitary waves 
from the expressions \eqref{width:r_near_1}, \eqref{height:r_near_1}, \eqref{impulse:r_near_1}, \eqref{energy:r_near_1}
for the width, relative-height, impulse, and energy in the slightly supersonic case. 
In terms of $r$ and $\V$, this yields
\begin{align}
\I_\s & \simeq \tfrac{2}{\sqrt{k-1}} \rho R \V (|\V|/c)^\frac{2}{k-1} \sqrt{1-r}
\label{impulse:weak}
\\
\E_\s & \simeq \tfrac{2(k+1)}{k\sqrt{k-1}} \rho R \V^2 (|\V|/c)^\frac{4}{k-1} \sqrt{1-r}  
\label{energy:weak}
\end{align}
for the impulse and the energy, respectively. 
Note both of these quantities scale like $\sqrt{1-r}$ when the wave speed is fixed. 

The width and the height of weakly nonlinear solitary waves 
are given in terms of $r$ and $\V$ by 
\begin{align}
\ell_\s & \simeq \tfrac{\twoS}{\sqrt{k-1}} R\big/\sqrt{1-r}
\label{width:weak}
\\
h_\s & \simeq (|\V|/c)^\frac{2}{k-1} (1-r) . 
\label{height:weak}
\end{align}
At any fixed value of $r$, 
the width and height have an interesting dependence on the Hertz exponent $k$. 
Specifically, the width $\ell_\s$ decreases with increasing $k$ 
and goes to $0$, 
while the height $h_\s$ also decreases and approaches the limit $1-r \neq 0$. 

More importantly, at any fixed wave speed $|\V|$,
the width scales like $1/\sqrt{1-r}$ in comparison to the particle size $2R$, 
while the height scales like $1-r$ in comparison to the background strain $v_0$. 
The same scaling proportionality between width and height 
is well-known to occur for solitons of the KdV equation,
which have a sech-squared profile. 
In fact, the KdV equation emerges in general for evolution systems that exhibit weak nonlinearity and dispersion \cite{Calagero}. 

We will now show that solitary waves \eqref{phys_solitarywave} in the weakly nonlinear regime 
approximately have the form of KdV solitons:
\begin{equation}\label{kdvsoln}
v/v_0 - 1 \simeq \tfrac{6}{k-1}(|\V|/c_0-1)\sech^2\Big( \sqrt{\tfrac{3}{2}}\sqrt{|\V|/c_0-1}\, (x-\V t)/R\Big)
\end{equation}
where  
\begin{equation}\label{kdvregime}
|\V|/c_0-1\simeq \tfrac{k-1}{6}(1-r) \ll 1 .
\end{equation}
Thus, the scaled wave profile as function of the scaled travelling wave variable is simply 
$(v-v_0)/h_\s \simeq \sech^2(\S|\zeta|/\ell_\s)$
where
\begin{equation}\label{kdvheight}
h_\s\simeq h_\kdv = \tfrac{6}{k-1}(|\V|/c_0-1) 
\end{equation}
is the height of the KdV soliton. 
The approximation here holds to leading order in $1-r$ uniformly in $v$. 

\begin{center}
\begin{figure}[!h]
\centering
\includegraphics[width=1.08\textwidth]{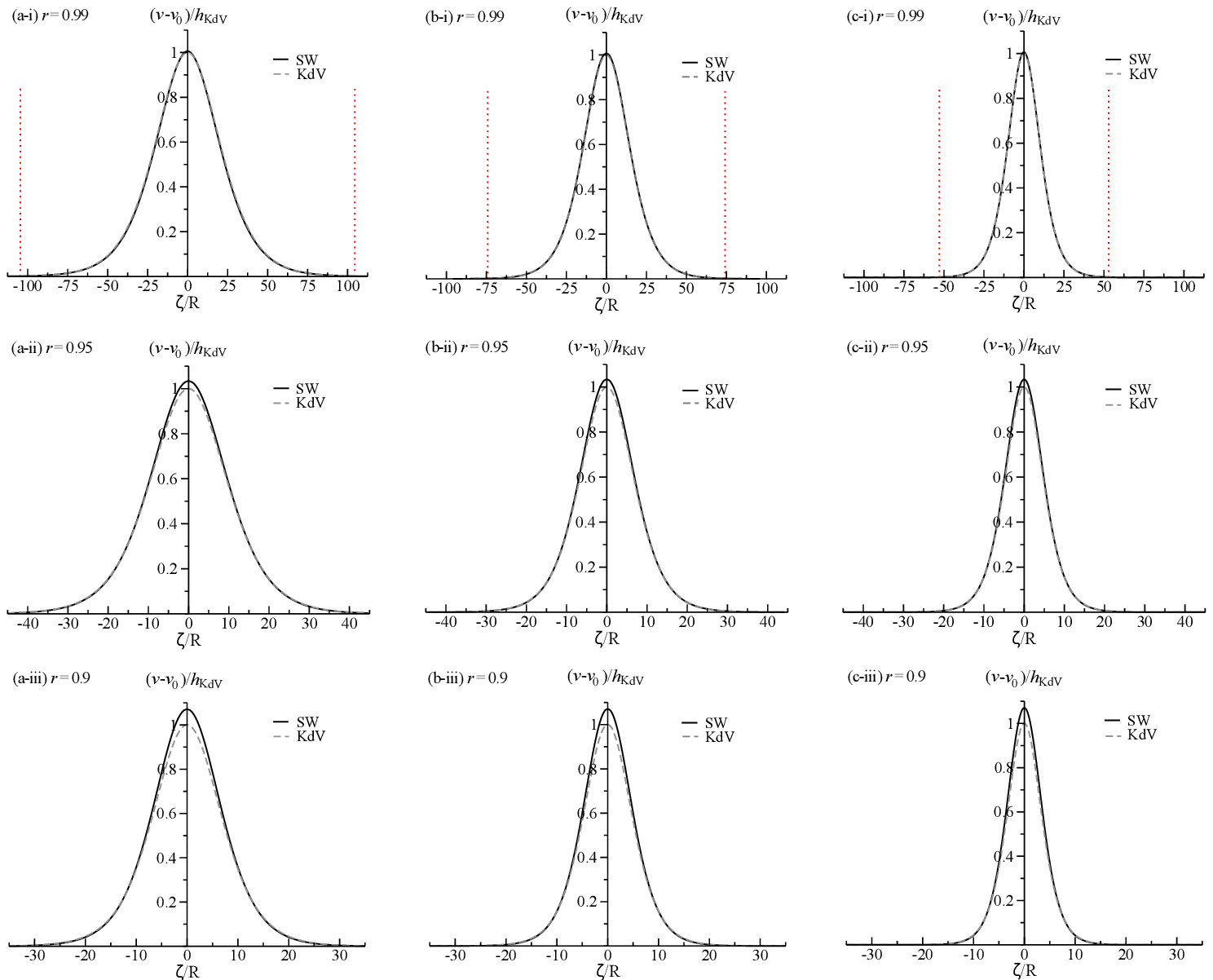}
\caption{
(Color online)
Comparison between KdV solitons \eqref{kdvsoln} and exact solitary wave (SW) solutions \eqref{phys_solitarywave} for $1-r \ll1$. 
Column (a) $k=3/2$; 
column (b) $k=2$; 
column (c) $k=3$. 
The solitary wave width is indicated by dashed vertical lines in row (i). 
\Eqref{kdvheight} for the height $h_\kdv$ is used to normalize both the KdV solitons and the exact solitary wave solutions.  
}
\label{fig:weakly_nonlinear_approx_plots}
\end{figure}
\end{center}

Fig.~\ref{fig:weakly_nonlinear_approx_plots} shows the KdV soliton approximation,
along with the corresponding exact solitary wave solutions from Section~\ref{sec:solitary-examples} for $1-r \ll 1$. 
It is evident that when $r$ is very close to $1$, 
the KdV soliton well-approximates the exact solitary wave solution for each value of $k$. 
The approximation worsens, particularly near the peak of the wave, 
as $r$ deviates further from $1$. 
In contrast to the quality of the approximations for strongly nonlinear solitary waves, 
which worsened with decreasing $k$ 
(cf.\ Figs.~\ref{fig:compacton_comparison_plots} and~\ref{fig:strongly_nonlinear_approx_plots}),
the quality of the normalized KdV approximation is roughly independent of $k$, 
for a given value of $r$. 

This approximation result is a continuum counterpart of what occurs \cite{JamesPelinovsky,YanLiuYanZhaDuaYan} 
in the weak nonlinearity limit for pre-compressed discrete chains \eqref{discrete-eom}. 
It is known that a two-scale expansion of the discrete equations of motion 
yields a discrete version of the KdV equation. 
We show in Appendix~\ref{appendix:KdV} that the same expansion 
can be applied to the LWHC wave equation \eqref{LWCE-u} to obtain the continuum KdV equation. 

To derive expression \eqref{kdvsoln}, 
we asymptotically expand the quadrature \eqref{phys_solitarywave}
for solitary waves 
\begin{equation}\label{smallv}
v=v(\zeta)= v_0(1 + w(\zeta))
\end{equation} 
such that $0<w\ll 1$ in the weakly nonlinear regime $1-r \ll 1$. 
In terms of $w$, this quadrature is given by 
\begin{equation}\label{solitarywave_smallv}
\int_{w}^{1/r-1} \frac{\sqrt{(1+w)^{k-1}}\, \,dw}{w\sqrt{\hat A(1+w,1;|\V|/c_0)}}  
= \tfrac{\sqrt{6}}{\sqrt{k(k+1)}} |\zeta|/R
\end{equation}
where
\begin{equation}
\hat A(1+w,1;|\V|/c_0) = 1- \tfrac{2}{k(k+1)}(c_0/\V)^2 (k-(1+w)\partial_w)\Big(\frac{(1+w)^k-1}{w}\Big) 
\end{equation}
from expression \eqref{hatA}. 
Expanding in powers of $w$ to leading non-trivial order, we get
$(1+w)^{k-1}/\hat A(1+w,1;|\V|/c_0) \simeq 1/(B_0 - B_1w)$
where
$B_0 = \tfrac{k(k+1)}{2}((|\V|/c_0)^2-1)$
and $B_1 =  \tfrac{k(k^2-1)}{6}((|\V|/c_0)^2-\tfrac{2}{3})$.
Then we use the asymptotic expansion \eqref{speedratio:weak} of the speed ratio $|\V|/c_0$ 
in terms of $1-r$, which yields
$B_0 - B_1w \simeq \tfrac{k(k^2-1)}{6}(1-r -w)$
and hence 
\begin{equation}
\frac{(1+w)^{k-1}}{\hat A(1+w,1;|\V|/c_0)} \simeq \tfrac{6}{k(k^2-1)} (1-r -w)^{-1} . 
\end{equation}
Thus, the asymptotic expansion of the quadrature \eqref{solitarywave_smallv} to leading order is given by 
\begin{equation}
\int_{w}^{1-r} \frac{dw}{w\sqrt{1-r -w}} 
\simeq \sqrt{k-1} |\zeta|/R . 
\end{equation}
This integral yields 
$\tfrac{2}{\sqrt{1-r}}\arctanh(\sqrt{1-r-w}/\sqrt{1-r})$,
and thus we obtain 
\begin{equation}
w \simeq (1-r)\sech^2\big(\tfrac{1}{2}\sqrt{1-r}\sqrt{k-1} |\zeta|/R\big) 
\end{equation}
which is valid for $1-r \ll 1$. 
The resulting solitary wave 
\begin{equation}
v=v_0\left(1 + (1-r)\sech^2\big(\tfrac{1}{2}\sqrt{1-r}\sqrt{k-1} |\zeta|/R\big)\right) 
\end{equation}
yields expression \eqref{kdvsoln}
when $1-r$ is expressed back in terms of the speed ratio \eqref{speedratio:weak}. 

\subsection{Comparisons and scaling relations}

The dependence of the width, height, impulse, and energy of solitary waves 
on the wave speed $\V$ is the same in both nonlinearity regimes. 
Specifically, when $r$ is fixed in each regime, 
the width is independent of the speed,
whereas the height scales like $h_\s\propto |\V|^\frac{2}{k-1}$
while the impulse and energy scale like 
$\I_\s\propto |\V|^\frac{k+1}{k-1}$ and $\E_\s\propto |\V|^\frac{2k+2}{k-1}$. 
Consequently, 
the latter three quantities satisfy the proportionality relationships:
\begin{equation}\label{solitary_rels1}
\I_\s \propto h_\s \V,
\quad
\E_\s \propto h_\s^2 \V^2
\end{equation}
and thus 
\begin{equation}\label{solitary_rels2}
\E_\s/\I_\s \propto h_\s \V \propto \sgn(\V) |\V|^\frac{k+1}{k-1},
\quad
\E_\s \propto \I_\s^2. 
\end{equation}

The same proportionality between energy and impulse holds 
for pre-compressed discrete chains \eqref{discrete-eom}
when compressive waves are generated by sharply striking an end particle in the chain, as noted in Section~\ref{model}.
Moreover, 
the two scaling relations $h_\s\propto |\V|^\frac{2}{k-1}$ and $\I_\s\propto |\V|^\frac{k+1}{k-1}$ 
agree with the ones reported in the literature \cite{Nesterenko2001,Daraio2006,Porter2008,Sen2008}
for weakly pre-compressed discrete chains. 

Finally, we note that the proportionality $\I_\s\propto |\V|^\frac{k+1}{k-1}$
indicates that all of the results on the properties of solitary waves 
in the weakly nonlinear and strongly nonlinear regimes 
can be expressed by replacing $\V$ in terms of $\I_\s$, 
using the respective impulse expressions \eqref{impulse:weak} and \eqref{impulse:r_gtr_1}. 
From a physical viewpoint, 
if we impart a specified sharp impulse to a pre-compressed continuum system,
then in either the weakly or strongly nonlinear regimes 
this impulse determines a unique solitary wave 
whose speed and height are given by 
$|\V| \propto |\I_\s|^\frac{k-1}{k+1}$, $h_\s \propto |\I_\s|^\frac{2}{k+1}$. 

It is more difficult to study solitary waves in the intermediate regime, 
where $r$ is neither close to $0$ nor close to $1$. 
By using expression \eqref{c0_r} for the sound speed in terms of $\V$ and $r$,
combined with expressions \eqref{height}, \eqref{width}, \eqref{impulse} and \eqref{energy} 
for the height $h_\s$, width $\ell_\s$, impulse $\I_\s$ and energy $\E_\s$, 
we obtain exact formulas that hold in any nonlinearity regime:
\begin{align}
& \ell_\s = 2R\Big( \sqrt{\tfrac{k(k+1)}{6}} \int_{1}^{1/r} \frac{\hat F(z,1;\mu(r))}{z-1} \,dz +\frac{\sqrt{3}}{\sqrt{\mu(r)^2 -1}} \Big) , 
%%% uses \S
\\
& h_\s = \big(\tfrac{|\V|}{c}\mu(r)\big)^{\frac{2}{k-1}}(1/r-1) ,
\\
& \I_\s = \tfrac{2}{\sqrt{3}} \rho Rc\, \sgn(\V) \bigg( \big(\tfrac{|\V|}{c}\mu(r)\big)^\frac{k+1}{k-1} \int_{1}^{1/r} \frac{\sqrt{z^{k-1}}}{\sqrt{\hat A(z,1;\mu(r))}}\,dz \bigg) , 
\\
&\begin{aligned}
\E_\s & = \tfrac{1}{\sqrt{3}} \rho Rc|\V|\bigg( \big(\tfrac{|\V|}{c}\mu(r)\big)^\frac{k+3}{k-1}  \int_{1}^{1/r} \sqrt{z^{k-1}}\bigg( \frac{z^2-1 + \tfrac{2}{k(k+1)}(1/\mu(r))^2( z^{k+1}-1 )}{(z-1)\sqrt{\hat A(z,1;\mu(r))}} 
\\&\qquad\qquad
- (z-1)\sqrt{\hat A(z,1;\mu(r))} \bigg)\,dz  \bigg) , 
\end{aligned}
\end{align}
where 
\begin{equation}
\mu(r) =  \sqrt{\tfrac{2}{k(k+1)}} \frac{\sqrt{r^{1-k} -k(1-r)r-r}}{1-r} . 
\end{equation}
These formulas do not exhibit any general scaling relations. 
Nevertheless, they can be evaluated straightforwardly for any $0<r<1$. 
Likewise, from the quadrature \eqref{phys_solitarywave} for solitary waves,
we have the exact integral formula for $v(x-\V t)$:
\begin{equation}\label{quadrature-r}
\int_{v/v_0}^{1/r} \frac{\sqrt{z^{k-1}}}{(z-1)\sqrt{\hat A(z,1;\mu(r))}} \,dz = \frac{\sqrt{3}\mu(r)}{R}|\zeta|,
\quad 
\zeta = x-\V t, 
\end{equation}
where
\begin{equation}
v_0 = \Big( \frac{|\V|/c}{\mu(r)} \Big)^\frac{2}{k-1} 
\end{equation} 
is the background strain. 
The transition between regimes of weak nonlinearity and strong nonlinearity 
can be seen in Fig.~\ref{fig:solution_plots}, 
using the explicit evaluation of this integral \eqref{quadrature-r}
for $k=3/2,2,3$ presented in Section~\ref{sec:solitary-examples}, 
where $r=g_0/g_1$ with $g_1$ given by the peak value of the scaled amplitude 
$g(\xi)=(g_0/v_0)v(\zeta)$. 

We see how the (scaled) profile of the solitary waves makes a continuous transition 
from a compacton-like shape when $r=0.001$ 
($g_0=0.001$, $g_1=0.999$) is small, 
to a KdV-soliton-like shape when $r=0.5$ 
($g_0=0.25$, $g_1=0.5$ for $k=2$; $g_0=0.3$, $g_1=0.6$ for $k=3$)
and $r=0.43$ ($g_0=0.2$, $g_1=0.47$ for $k=3/2$). 

An analysis of the continuous transition from weak nonlinearity to strong nonlinearity for solitary waves has not previously appeared in the literature. 
We remark that the nonlinearity transition for shock waves in discrete systems 
have been studied in \Ref{GomTurHecVit,GomTurVit}.

\section{Concluding remarks}\label{remarks}

In this paper we have presented a physical analysis of 
the properties of long wavelength solitary waves 
in the continuum model of Hertzian chains with arbitrary pre-compression. 

We find that the ratio of the background strain to the peak strain 
in solitary wave solutions describes the degree of dynamical nonlinearity in the underlying discrete chain. 
This dynamical nonlinearity ratio is determined by a nonlinear algebraic relation given in terms of the ratio of the solitary wave speed to the sound speed. 
In particular, 
highly supersonic solitary waves correspond to highly nonlinear propagating localized pulses in weakly compressed discrete chains, 
while slightly supersonic solitary waves correspond to weakly nonlinear propagating localized pulses in strongly compressed discrete chains. 

We have obtained explicit analytical expressions for the height, width, impulse and energy of solitary waves. 
The width expression is a new result, coming from an asymptotic analysis of the tail of solitary waves. 
Using this expression, 
we have shown that the width of all solitary waves 
decreases with $k$ at any fixed value of the nonlinearity ratio, 
and increases with the nonlinearity ratio, at any fixed value of $k>1$. 
In the physically interesting case when $k=3/2$ (corresponding to spherical particles), 
the minimum width is approximately $5$ times the particle size. 
This confirms the validity of the continuum model for studying solitary waves. 

We have used the height, width, impulse and energy expressions to show that the main physical features of solitary waves depend principally on the ratio of the wave speed to the sound speed at any fixed $k>1$. 
Moreover, highly supersonic solitary waves are shown to be well-approximated
by Nesterenko's compacton,
while slightly supersonic solitary waves are shown to coincide approximately 
with KdV solitons which have a well-known sech-squared profile. 
We further have shown that the KdV equation itself arises directly from the LWHC equation through a two-scale asymptotic expansion 
combined with a Galilean transformation to a reference frame moving with the sound speed. 

Exact solitary wave solutions have been used to compare the features of 
solitary waves across different nonlinearity regimes. 
Specifically, we have derived an exact expression for the solitary waves that 
arise in the continuum model with a Hertz exponent $k=3/2$, 
corresponding to solitary wave pulses in a chain of spherical discrete particles. 
The shape of these solitary waves is highly sensitive to their speed. 

Our results establish that a continuum system supports 
solitary waves having the same impulse momentum 
yet displaying marked different shapes and speeds, 
especially in comparison to a compacton. 

The same conclusion will hold for physical discrete chains 
when the long wavelength regime is considered. 

Our long wavelength continuum analysis can be extended to models of heterogeneous chains, in particular dimer chains consisting of alternating particles with different masses. 
There are several further interesting directions for future work. 

One direction would be to study the properties of solitary waves in a finite-size continuum model with physical boundary conditions. 
This will yield analytical results concerning the reflection of solitary waves at the end point boundaries,
which is important for understanding physical discrete chains. 

Another direction would be to investigate the statistical properties of a dilute (rarified) ensemble of interacting solitary waves, known as a soliton gas. 
This is relevant for the dynamics of discrete chains after transients have decayed such that the chain contains propagating solitary waves with a wide spectrum of energies. 
Of notable interest here is the recent observation that rogue waves can form in such circumstances \cite{SluPel,ShuPel,PelShu}.

\section*{Acknowledgments}
S.C.A.\ is supported by an NSERC research grant.
The work of M.P.\ was supported by a Vanier Canada Graduate Scholarship.

\appendix

\section{Derivation of $k=\tfrac{3}{2}$ solitary wave solution}\label{appendix:k=3/2}

Here we outline the main steps in the derivation of the solitary wave solution \eqref{solitarysol3}
for $k=\tfrac{3}{2}$,
starting from the quadrature \eqref{solitary_ode_integral},
which is given by 
\begin{equation}\label{solitary_ode_integral_2k=3}
\pm \int_{g}^{g_1} \frac{g^{1/4}}{\sqrt{(g-g_0)^2+\tfrac{3}{2}g_0^{3/2}(g-g_0)-g(g^{3/2}-g_0^{3/2})}} \,dg = |\xi| .
\end{equation}

As shown in \Ref{PrzAnc}, 
the change of variable $g=h^2$ brings this integral \eqref{solitary_ode_integral_2k=3} 
to the form of an elliptic integral 
\begin{equation}\label{solitary_elliptic_integral_2k=3}
\pm 2 \int_{h}^{h_1} \frac{h^2}{(h-h_0)\sqrt{B(h)}} \,dh = |\xi| 
\end{equation}
where 
\begin{equation}
B(h)=h((1-\tfrac{3}{2}h_0)h_0^2+(2-3h_0)h_0h+(1-2h_0)h^2 -h^3)
\end{equation}
is a quartic polynomial. 
It is straightforward to show that this polynomial has the factorization
$B(h)=B_1(h)B_2(h)$ with $B_1(h)=h(h_1-h)$ and $B_2(h)=(h+h_2)^2+h_3^2$, 
where $h_1,h_2,h_3$ are given by expressions \eqref{h1-expr}--\eqref{h2h3-expr}. 
Standard methods \cite{Abramowitz1964,Lawden1980} can now be used to evaluate
the elliptic integral \eqref{solitary_elliptic_integral_2k=3} explicitly. 

The first step is to change variables by applying a linear fractional transformation 
\begin{equation}\label{eq:y}
y= (h-y_+)/(h-y_-),
\end{equation} 
with $y_\pm$ being defined by the condition 
$B_1(h) -\lambda_\pm B_2(h)= -(1+\lambda_\pm)(h-y_\pm)^2$,
where $\lambda_\pm$ are the roots of the discriminant of $B_1(h)-\lambda B_2(h)$. 
This leads to the expressions 
\begin{equation}\label{eq:BYrel}
(y-1)^2R B_1(h) = \sqrt{PQ} Y_1(y),
\quad
(y-1)^2R B_2(h) = \sqrt{PQ} Y_2(y)
\end{equation} 
given in terms of the quadratic polynomials
\begin{equation}
Y_1(y) = K_-^2 - K_+^2 y^2,
\quad
Y_2(y) = (K_+^2 -R)y^2 +R- K_-^2.
\end{equation}
where
\begin{equation}
K_{\pm} = \sqrt{Q}\pm\sqrt{P}
\end{equation}
and
\begin{equation}
P=h_2^2+h_3^2=B_2(0),
\quad
Q=(h_1+h_2)^2+h_3^2=B_2(h_1),
\quad
R=(h_1+2h_2)^2.
\end{equation}
Carrying out the change of variable \eqref{eq:y},
and splitting up the integral into terms with even/odd parity under $y\rightarrow -y$, 
we get 
\begin{equation}\label{eq:integralterms}
\int_{h}^{h_1} \frac{h^2\, dh}{(h-h_0)\sqrt{B(h)}} 
= c_1 I_1+ c_2 I_2 +c_0 J_0 +c_1 J_1 +c_2 y_0 J_2,
\end{equation} 
where
\begin{align}
& 
I_1=\int_{y}^{y_1} \frac{y\,dy}{(y^2-1)\sqrt{Y_1(y)Y_2(y)}}, 
\
I_2 = \int_{y}^{y_1} \frac{y\,dy}{(y^2-y_0^2)\sqrt{Y_1(y)Y_2(y)}}, 
\label{eq:integral-I}
\\
& 
J_0 = \int_{y}^{y_1} \frac{dy}{\sqrt{Y_1(y)Y_2(y)}}, 
\ 
J_1 = \int_{y}^{y_1} \frac{dy}{(y^2-1)\sqrt{Y_1(y)Y_2(y)}}, 
\ 
J_2 = \int_{y}^{y_1} \frac{dy}{(y^2-y_0^2)\sqrt{Y_1(y)Y_2(y)}}
\label{eq:integral-J}
\end{align}
and
\begin{equation}
c_0= \frac{y_-^2(y_0-1)R}{\sqrt{PQ}},
\quad
c_1= -\frac{(y_+-y_-)^2R}{\sqrt{PQ}},
\quad
c_2= \frac{h_0^2(y_0-1)^2R}{\sqrt{PQ}},
\label{eq:c1c2c3}
\end{equation}
with 
\begin{equation}\label{eq:ypm}
y_\pm = \sqrt{P}(\pm\sqrt{Q} -\sqrt{P})/\sqrt{R}
\end{equation} 
and
\begin{align}
& y_1 = (h_1-y_+)/(h_1-y_-) = K_-/K_+,
\\
& y_0 = (h_0-y_+)/(h_0-y_-) = \frac{h_0\sqrt{R}-K_-\sqrt{P}}{h_0\sqrt{R}+K_+\sqrt{P}}.
\end{align}
We note that the integral \eqref{eq:integralterms} must diverge to $\infty$ when $h\to h_0$,
since this limit corresponds to the tail of the solitary wave $|\xi|\to \infty$. 
In terms of the variable $y$, this divergence occurs for $y\to y_0$. 

The five separate integrals \eqref{eq:integral-I}--\eqref{eq:integral-J} 
can be simplified by a further change of variables. 
We introduce 
\begin{equation}\label{eq:z}
z=y/y_1,
\end{equation}
and let 
\begin{gather}
\tau= K_-\sqrt{R- K_-^2},
\\
k= y_1^2(K_+^2-R)/(R-K_-^2),
\quad
n=(y_1/y_0)^2,
\quad
m=y_1^2.
\end{gather}
Note $y\rightarrow y_0$ now corresponds to $z\rightarrow y_0/y_1 =1/\sqrt{n}$. 

The first two integrals \eqref{eq:integral-I} 
can be evaluated directly in terms of elementary functions:
\begin{equation}\label{eq:I1}
\begin{aligned}
\frac{\tau}{y_1^2} I_1 & 
= \int_{z}^{1} \frac{z\, dz}{(y_1^2z^2-1)\sqrt{(1-z^2)(1+kz^2)}} 
\\& 
= \frac{-1}{2\sqrt{(1-m)(k+m)}}
\bigg( \frac{\pi}{2} + \arctan\bigg( \frac{(m-1)(1+kz^2) +(k+m)(1-z^2)}{2\sqrt{(k+m)(1-m)}\sqrt{(1-z^2)(1+kz^2)}} \bigg) \bigg), 
\end{aligned}
\end{equation}
and 
\begin{equation}\label{eq:I2}
\begin{aligned}
\frac{y_0^2\tau}{y_1^2} I_2 & 
= \int_{z}^{1} \frac{z\,dz}{((y_1/y_0)^2z^2-1)\sqrt{(1-z^2)(1+kz^2)}} 
\\& 
= \frac{1}{2\sqrt{(n-1)(k+n)}}
\ln\bigg( \frac{\big(\sqrt{n-1} \sqrt{1+kz^2} +\sqrt{k+n} \sqrt{1-z^2}\big)^2}{(k+1)(nz^2-1)} \bigg). 
\end{aligned}
\end{equation}
For $z\rightarrow y_0/y_1 =1/\sqrt{n}$, 
the integral $I_1$ is finite, 
while the integral $I_2$ has a logarithmic singularity 
$(y_0/y_1)^2\tau I_2 \to \tfrac{-1}{2\sqrt{(n-1)(k+n)}} \ln(nz^2-1) \to +\infty$. 

The remaining three integrals \eqref{eq:integral-J}
can be evaluated in terms of Jacobi elliptic functions:
\begin{align}
&
\tfrac{\tau}{y_1} J_0 
= \int_{z}^{1} \frac{dz}{\sqrt{(1-z^2)(1+kz^2)}} 
= \tfrac{1}{\sqrt{1+k}} \cn^{-1}(z|l),
\label{eq:J0}
\\
&
\tfrac{\tau}{y_1} J_1 
=\int_{z}^{1} \frac{dz}{(y_1^2z^2-1)\sqrt{(1-z^2)(1+kz^2)}} 
= \tfrac{1}{(m-1)\sqrt{1+k}} \Pi\big(\tfrac{m}{m-1};\cn^{-1}(z|l)|l\big),
\label{eq:J1}
\\
&
\tfrac{\tau y_0^2}{y_1} J_2 
= \int_{z}^{1} \frac{dz}{((y_1/y_0)^2z^2-1)\sqrt{(1-z^2)(1+kz^2)}} 
= \tfrac{1}{(n-1)\sqrt{k+1}} \Pi\big(\tfrac{n}{n-1};\cn^{-1}(z|l)|l\big),
\label{eq:J2}
\end{align}
with
\begin{equation}
l=k/(1+k).
\end{equation}
For $z\rightarrow y_0/y_1 =1/\sqrt{n}$, 
the integrals $J_0$ and $J_1$ are finite, 
while the integral $J_2$ can be shown to have a logarithmic singularity. 

The next step consists of extracting the singular part of $J_2$ 
by using the elliptic function identity~\cite{Abramowitz1964} 
\begin{equation}
\Pi\big(N;\theta|l\big)
= -\Pi\big(l/N;\theta|l\big) + \theta + \tfrac{1}{2}\psi\ln\Big(\frac{\psi +\sn(\theta|l)/(\cn(\theta|l)\dn(\theta|l))}{\psi -\sn(\theta|l)/(\cn(\theta|l)\dn(\theta|l))}\Big),
\quad 
N>1,
\end{equation}
where $\psi = \tfrac{1}{\sqrt{(N-1)(1-l/N)}}$. 
This gives
\begin{equation}\label{eq:logterm}
\begin{aligned}
\Pi\big(n/(n-1);\cn^{-1}(z|l)|l\big) & = 
\cn^{-1}(z|l) -\Pi\big(l(n-1)/n;\cn^{-1}(z|l)|l\big)
\\&\qquad
+ \tfrac{1}{2}\psi\ln\Big(\frac{\tilde\psi +\sqrt{1-z^2}/(z\sqrt{1+kz^2})}{\tilde\psi -\sqrt{1-z^2}/(z\sqrt{1+kz^2})}\Big)
\end{aligned}
\end{equation}
with $\tilde\psi = \frac{\psi}{\sqrt{k+1}}= \tfrac{\sqrt{1-1/n}}{(1/\sqrt{n})\sqrt{1+k/n}}$. 
In the logarithm term in expression \eqref{eq:logterm},
the denominator vanishes at $z^2=1/n$,
and so we can factorize
\begin{equation}
\begin{aligned}
\frac{\tilde\psi +\sqrt{1-z^2}/(z\sqrt{1+kz^2})}{\tilde\psi -\sqrt{1-z^2}/(z\sqrt{1+kz^2})} & =
\frac{(n-1)( z\sqrt{1+kz^2} + (1/\tilde\psi)\sqrt{1-z^2} )^2}{(nz^2-1)( 1+kz^2 + (k/n)(1-z^2) )}.
\end{aligned}
\end{equation}
Thus, we can write
\begin{equation}
J_2 = J_{2,0}+J_{2,1}+I_3,
\end{equation}
where
\begin{equation}\label{eq:J2-elliptic}
\frac{\tau y_0^2}{y_1} J_{2,0} 
= \tfrac{1}{(n-1)\sqrt{k+1}} \cn^{-1}(z|l),
\quad
\frac{\tau y_0^2}{y_1} J_{2,1} 
= \tfrac{-1}{(n-1)\sqrt{k+1}} \Pi\big(l(n-1)/n;\cn^{-1}(z|l)|l\big), 
\end{equation}
are finite for $z\rightarrow y_0/y_1 =1/\sqrt{n}$, 
and where
\begin{equation}\label{eq:J2-rat}
\frac{\tau y_0^2}{y_1} I_3 
= \tfrac{\sqrt{n}}{2\sqrt{k+n}} \ln\Big(\frac{(n-1)( z\sqrt{1+kz^2} + (1/\tilde\psi)\sqrt{1-z^2} )^2}{(nz^2-1)( 1+kz^2 + (k/n)(1-z^2) )}\Big). 
\end{equation}
has the logarithmic singularity 
$(y_0^2/y_1)\tau I_3 \to \tfrac{-\sqrt{n}}{2\sqrt{k+n}} \ln(nz^2-1) \to +\infty$. 

All of the integrals \eqref{eq:I1}, \eqref{eq:I2}, \eqref{eq:J0}, \eqref{eq:J1}, \eqref{eq:J2-elliptic}, and \eqref{eq:J2-rat}
can be written explicitly in terms of $h$ through the relations
\begin{equation}
z=y/y_1 = \frac{K_+ h-\sqrt{P} h_1}{K_- h +\sqrt{P} h_1}, 
\quad
Y_1(y)/Y_2(y)=B_1(h)/B_2(h)
\end{equation}
obtained by inverting the change of variables \eqref{eq:y} and \eqref{eq:z} 
and by using equations \eqref{eq:y}, \eqref{eq:BYrel}, \eqref{eq:ypm}, 
%\begin{equation}
%1-y = \frac{2\sqrt{PQ}}{\sqrt{R}h+\sqrt{P}K_+},
%y-y_0 = \frac{\sqrt{R}(1-y_0)}{2\sqrt{PQ}} (1-y)(h-h_0)
%1-y_0= \frac{2\sqrt{PQ}}{\sqrt{P}K_++h_0\sqrt{R}}.
%\end{equation}

For the final step, 
the combined elementary integrals $c_1I_1 +c_2(I_2 +y_0 I_3) =I(h)$ 
can be expressed in the form \eqref{solitarysol3-rat-terms}
after some simplifications using the following identities:
First, from relation \eqref{eq:BYrel} evaluated at $y=y_0$ and $y=y_1$, 
we have 
\begin{align}
& K_-^2 - K_+^2 y_0^2 
= \frac{(y_0-1)^2 h_0(h_1-h_0)R}{\sqrt{PQ}},
\label{eq:B1x0}\\
& (K_+^2 -R)y_0^2 +R- K_-^2 
= \frac{(y_0-1)^2 RS}{\sqrt{PQ}},
\label{eq:B2x0}\\
& (K_+^2 -R)y_1^2 +R- K_-^2 
= \frac{(y_1-1)^2 R\sqrt{Q}}{\sqrt{P}}.
\label{eq:B2x1}
\end{align}
Next, we find 
\begin{equation}
K_+K_- = h_1\sqrt{R},
\quad
K_++K_- = 2\sqrt{Q},
\quad
K_+-K_- = 2\sqrt{P},
\end{equation}
and
\begin{equation}
y_+-y_-=\frac{2\sqrt{PQ}}{\sqrt{R}},
\quad
1-y_1 = \frac{2\sqrt{P}}{K_+}.
\end{equation}
Last, we can derive 
\begin{gather}
1-m=\frac{4\sqrt{PQ}}{K_+^2},
\quad
k+m= \frac{4\sqrt{PQ}K_-^2}{R(K_+^2-h_1^2)},
\quad
k+1
%= \frac{R\sqrt{Q}(1-y_1)^2}{(R-K_-^2)\sqrt{P}}
= \frac{\sqrt{PQ}}{K_+^2-h_1^2},
\\
n-1= \frac{Rh_0(h_1-h_0)(1-1/y_0)^2}{K_+^2\sqrt{PQ}},
\quad
k+n 
%= \frac{RS y_1^2(1-1/y_0)^2}{(R-K_-^2)\sqrt{PQ}}
= \frac{S K_-^2}{(K_+^2-h_1^2)\sqrt{PQ}} 
\end{gather}
which yields
\begin{gather}
\frac{1-z^2}{1+kz^2} = \frac{K_+^2-h_1^2}{h_1^2} \frac{Y_1(y)}{Y_2(y)},
%\frac{R-K_-^2}{K_-^2} 
\quad
\frac{k+m}{1-m} = \frac{h_1^2}{K_+^2-h_1^2},
%\frac{K_-^2}{R-K_-^2}
\quad
\frac{k+n}{n-1} = \frac{h_1^2S}{(K_+^2-h_1^2)h_0(h_1-h_0)}. 
\end{gather}
Similarly, 
the combined elliptic integrals $c_0 J_0 +c_1 J_1 +c_2 y_0 (J_{2,0}+J_{2,1})=J(h)$
can be expressed in the form \eqref{solitary-sol3-cn_inverse} 
by using the relations
\begin{gather}
l = \frac{k}{k+1}
= \frac{h_1^2-K_-^2}{4\sqrt{PQ}},
\quad
\frac{m}{1-m} = \frac{K_-^2}{4\sqrt{PQ}},
\\
\frac{n}{n-1} = \frac{K_-^2\sqrt{PQ}}{Rh_0(h_1-h_0)(1-y_0)^2},
\quad
\frac{l(n-1)}{n} = \frac{(K_+^2-R)h_0(h_1-h_0)(1-y_0)^2}{4PQ}.
\end{gather}

\section{Asymptotic expansions of width, impulse, and energy}\label{appendix:expansion}

Here we explain the steps for asymptotically expanding 
the width integral \eqref{width_integral:r}, impulse integral \eqref{impulse_integral:r} and energy integral \eqref{energy_integral:r} 
in the slightly supersonic/weakly nonlinear and highly supersonic/strongly nonlinear cases.

As a first step, it is very helpful to make a change of integration variable 
$z=g/g_0$, where $g_0=v_0/\lambda$ is expression \eqref{g0}. 
By combining this expression and the sound speed expression \eqref{soundspeed},
we have the useful relations
\begin{equation}\label{speed_g:rel}
%g_0 =  (c_0/|\V|)^\frac{2}{k-1} g^*
|\V|/c_0 = (g^*/g_0)^\frac{k-1}{2} = (g^*\lambda)^\frac{k-1}{2} c/c_0 , 
\end{equation}
where $g^*$ is expression \eqref{g1} which involves only $k$. 
Then the integrals \eqref{width_integral:r}, \eqref{impulse_integral:r}, \eqref{energy_integral:r} 
are respectively given by 
\begin{align}
\hat\ell & = \sqrt{\tfrac{k(k+1)}{2}}\int_{g_0}^{g_1} \frac{F(g,g_0)}{g-g_0} \,dg + \S/\sqrt{(g_0/g^*)^{1-k} - 1} ,
\label{width:g}
\\
\hat\I & = \lambda^\frac{k+1}{2} \int_{g_0}^{g_1} \frac{\sqrt{g^{k-1}}}{\sqrt{A(g,g_0)}}\,dg,
\label{impulse:g}
\\
\hat\E & = \lambda^\frac{k+3}{2} \int_{g_0}^{g_1} \sqrt{g^{k-1}}\left( \frac{g^2-g_0^2 + g^{k+1}-g_0^{k+1} }{(g-g_0)\sqrt{A(g,g_0)}} - (g-g_0)\sqrt{A(g,g_0)} \right)\,dg . 
\label{energy:g}
\end{align}
%$\lambda =(\tfrac{k(k+1)}{2}\V^2/c^2)^\frac{1}{k-1}$

\subsection{Slightly supersonic/weakly nonlinear case}
This case $|\V|/c_0-1 \ll 1$ corresponds to $(g^*/g_0)^\frac{k-1}{2} -1 = \epsilon \ll 1$ 
from the first relation \eqref{speed_g:rel}. 
Hence we have $g_0 \simeq (1-\tfrac{2}{k-1}\epsilon)g^*$, 
and $g_1 = g_0/r \simeq (1+\tfrac{4}{k-1}\epsilon)g^*$ 
from the strain ratio \eqref{r_near1}. 
In particular, note $(g_1-g_0)/g^* \simeq \tfrac{6}{k-1}\epsilon$.

To asymptotically expand the integrals \eqref{width:g}, \eqref{impulse:g}, \eqref{energy:g} in terms of $\epsilon$, 
we first take the leading-order term in the integrand, 
which is given by substitution of $g=g^*$, 
and then we multiply this term by the endpoint difference 
$g_1-g_0 \simeq \tfrac{6}{k-1} g^*\epsilon$.

For the width integral \eqref{width:g}, 
this expansion yields 
\begin{equation}
\begin{aligned}
\hat\ell - \S/\sqrt{(g_0/g^*)^{1-k} - 1}
& \simeq \sqrt{\tfrac{k(k+1)}{2}}(g_1-g_0) \frac{F(g^*,g_0)}{g^*-g_0} \\
& \simeq 3\sqrt{\tfrac{k(k+1)}{2}}F(g^*,g_0) \\
\end{aligned}
\end{equation}
which is $O(\epsilon)$ since $F(g_0,g_0)=0$. 
But since $1/\sqrt{(g_0/g^*)^{1-k} - 1} \simeq 1/\sqrt{2\epsilon}$, 
the leading-order term in the width integral is given by 
$\hat\ell \simeq \S/\sqrt{2\epsilon}$.

Expansion of the impulse integral \eqref{impulse:g} directly yields
\begin{equation}
\hat\I \simeq \lambda^\frac{k+1}{2} (g_1-g_0) \sqrt{g^*{}^{k-1}/A(g^*,g_0)} \\
\simeq \tfrac{6}{k-1} \big( (g^*\lambda)^\frac{k+1}{2} /\sqrt{A(g_0,g_0)} \big)\epsilon
\end{equation}
where, from expression \eqref{A_props}, $A(g_0,g_0)\simeq 2\epsilon$.
Moreover, we note $g^*\lambda \simeq (c_0/c)^\frac{2}{k-1}$ from the relations \eqref{speed_g:rel}. 
Hence the leading-order term in the impulse integral is given by 
$\hat\I \simeq \tfrac{3\sqrt{2}}{k-1} (c_0/c)^\frac{k+1}{k-1} \sqrt{\epsilon}$.

Similarly, the energy integral \eqref{energy:g} has the expansion
\begin{equation}
\begin{aligned}
\hat\E  & \simeq \lambda^\frac{k+3}{2} (g_1-g_0) \sqrt{g^*{}^{k-1}}\left( \frac{g^*{}^2-g_0^2 + g^*{}^{k+1}-g_0^{k+1} }{(g^*-g_0)\sqrt{A(g^*,g_0)}} - (g^*-g_0)\sqrt{A(g^*,g_0)} \right) \\
& \simeq \tfrac{6}{k-1} \lambda^\frac{k+3}{2} \sqrt{g^*{}^{k+3}/A(g^*,g_0)} (2 + (k+1) g^*{}^{k-1}) \epsilon \\
& \simeq \tfrac{12}{k-1} (1 + \tfrac{1}{k}) \big( (g^*\lambda)^\frac{k+3}{2}\big/\sqrt{A(g_0,g_0)}\big) \epsilon . 
\end{aligned} 
\end{equation}
Hence the leading-order term is given by 
$\hat\E \simeq \tfrac{6\sqrt{2}(k+1)}{k(k-1)} (c_0/c)^\frac{k+3}{k-1} \sqrt{\epsilon}$.

\subsection{Highly supersonic/strongly nonlinear case}
This case $|\V|/c_0 \gg 1$ corresponds to $(g_0/g^*)^\frac{k-1}{2} =\epsilon \ll 1$
from the first relation \eqref{speed_g:rel}. 
Hence we have 
$g_0 = \epsilon^\frac{2}{k-1} g^*$ 
and $g_1 = g_0/r \simeq 1$ from the strain ratio \eqref{r_near0}. 
For the subsequent steps, 
it will be helpful to note 
\begin{equation}\label{g1g0rel}
1-g_1 \simeq \tfrac{2}{k-1}g_0 .
\end{equation}
This is obtained by solving equation \eqref{rroot} to second order in powers of $\epsilon^\frac{2}{k-1}$. 

To asymptotically expand the impulse integral \eqref{impulse:g} and energy integral \eqref{energy:g},
we use a Taylor series in $g_0$ and explicitly evaluate the first two terms
\begin{align}
\hat\I & = \hat\I|_{g_0=0} + g_0\partial_{g_0}\hat\I|_{g_0=0} + O(g_0^2),
\\
\hat\E & = \hat\E|_{g_0=0} + g_0\partial_{g_0}\hat\E|_{g_0=0} + O(g_0^2) .
\end{align}
Since $g_0 =O(\epsilon^\frac{2}{k-1})$, the first term in the Taylor series gives
the leading order term for the asymptotic expansions,
while the second term gives the subleading term. 

The leading-order term in each expansion is given by evaluating
\begin{equation}
\lambda^{-\frac{k+1}{2}} \hat\I|_{g_0=0} 
%= \lambda^\frac{k+1}{2} \int_{g_0}^{g_1} \frac{\sqrt{g^{k-1}}}{\sqrt{A(g,g_0)}}\,dg
= \int_{0}^{1} \frac{\sqrt{g^{k-1}}}{\sqrt{A(g,0)}}\,dg 
= \int_{0}^{1} \frac{dg}{\sqrt{g^{1-k}-1}}
\end{equation}
and
\begin{equation}
\begin{aligned}
\lambda^{-\frac{k+3}{2}} \hat\E|_{g_0=0}
% = \lambda^\frac{k+3}{2} \int_{g_0}^{g_1} \sqrt{g^{k-1}}\left( \frac{g^2-g_0^2 + g^{k+1}-g_0^{k+1} }{(g-g_0)\sqrt{A(g,g_0)}} - (g-g_0)\sqrt{A(g,g_0)} \right)\,dg
= \int_{0}^{1} \sqrt{g^{k+1}}\left( \frac{1 + g^{k-1}}{\sqrt{A(g,0)}} - \sqrt{A(g,0)} \right)\,dg 
= 2 \int_{0}^{1} \frac{g^{k}\,dg}{\sqrt{g^{1-k}-1}} ,
\end{aligned}
\end{equation}
where $A(g,0)=1-g^{k-1}$ from expression \eqref{A}. 
This yields
\begin{align}
\hat\I & = 
\lambda^\frac{k+1}{2} \sqrt{\pi} \Gamma\big(\tfrac{k+1}{2k-2}\big)/\Gamma\big(\tfrac{1}{k-1}\big)
+O(\epsilon^\frac{2}{k-1}),
\label{impulse:leadingterm}
\\
\hat\E & = 
\lambda^\frac{k+3}{2} \sqrt{\pi} \tfrac{k+3}{4k+4}\Gamma(\tfrac{k+3}{2k-2})/\Gamma(\tfrac{2}{k-1}) 
+O(\epsilon^\frac{2}{k-1}) .
\label{energy:leadingterm}
\end{align}

For the width integral \eqref{width:g},
we separately consider the algebraic term and the integral term. 
The algebraic term can be expanded in a Taylor series in $(g_0/g^*)^\frac{k-1}{2} =\epsilon$, 
which yields
\begin{equation}
\S/\sqrt{(g_0/g^*)^{1-k} - 1} = \S\epsilon  + O(\epsilon^2) . 
\end{equation}
In contrast, the integral term is more complicated to analyse, 
and compared to the impulse integral and the energy integral, 
it does not possess a Taylor series in $g_0$ beyond the leading order term. 
To obtain an asymptotic expansion, 
we will first split up the integrand into a term given by the limit $g_0\to 0$
and a remainder term given by subtracting off this limit term. 

The limit $g_0\to 0$ of the integrand $F(g,g_0)/(g-g_0)$ yields
$F(g,0)/g = \sqrt{g^{k-3}/(1-g^{k-1})}$ through expression \eqref{F}. 
Hence we split up the integral 
\begin{equation}\label{r_near0:scaledwidth_integral}
\int_{g_0}^{g_1} \frac{F(g,g_0)}{g-g_0} \,dg
= \int_{g_0}^{g_1} \frac{F_1(g)}{g} \,dg 
+ \int_{g_0}^{g_1} \frac{g_0F_1(g)+gF_2(g,g_0)}{g(g-g_0)} \,dg
\end{equation}
where
\begin{equation}
\quad
F_1(g)=F(g,0),
\quad
F_2(g,g_0) = F(g,g_0)-F_1(g) . 
\end{equation}
The first integral term on the right-hand side of equation \eqref{r_near0:scaledwidth_integral}
can be evaluated explicitly
\begin{equation}
\int_{g_0}^{g_1} \frac{F(g,0)}{g} \,dg
= \int_{g_0}^{g_1} \frac{\sqrt{g^{k-3}}}{\sqrt{1-g^{k-1}}} \,dg
= \tfrac{2}{k-1} \big( \arcsin(g_1{}^\frac{k-1}{2}) - \arcsin(g_0{}^\frac{k-1}{2}) \big) ,
\end{equation}
which gives 
\begin{equation}
\int_{g_0}^{g_1} \frac{F(g,0)}{g} \,dg \simeq \tfrac{\pi}{k-1} +O(\epsilon^\frac{1}{k-1},\epsilon) . 
\end{equation}
To analyse the remaining integral term on the right-hand side of equation \eqref{r_near0:scaledwidth_integral}, 
we need to take into account that the integrand has a square-root singularity 
because 
\begin{equation}
A(g,g_0)=(g_1-g)B(g,g_0), 
\quad
B(g_1,g_0)= \frac{(k+1)(g_1^k-g_0^k)}{(g_1-g_0)^2} \neq 0,
\end{equation}
and we also need to note that, through relation \eqref{g1g0rel}, 
$F_1(g_1)\simeq g_1{}^\frac{k-1}{2}/\sqrt{2g_0}$ is singular when $g_0\to 0$. 
Consequently, we first combine these singular terms by expressing 
\begin{equation}
\begin{aligned}
g_0F_1(g)+gF_2(g,g_0) =& 
\frac{(g-g_0)C(g,g_0)}{\sqrt{g_1-g}\sqrt{B(g,g_0)}\big(g_0{}^\frac{k-1}{2}g\sqrt{A(g,g_0)} + g_0\sqrt{A(g_0,g_0)} g^\frac{k-1}{2}\big)}
\\&
+\frac{g^\frac{k-1}{2}(g-g_0)H(g,g_0)}{\sqrt{g_1-g}\sqrt{B(g,g_0)}\big(\sqrt{1-g^{k-1}} + \sqrt{A(g,g_0)}\big)}
\end{aligned}
\end{equation}
where
\begin{equation}\label{r_near0:H}
H(g,g_0) = \sqrt{1-g^{k-1}} -\frac{A(g,g_0)}{\sqrt{1-g^{k-1}}}
\end{equation}
and
\begin{equation}\label{r_near0:C}
C(g,g_0) = g_0^{k-1}\frac{(g+g_0)A(g,g_0)}{\sqrt{A(g_0,g_0)}} -
g_0^2\sqrt{A(g_0,g_0)}\Big( \frac{g^{k-1}-g_0^{k-1}}{g-g_0} + g_0^{k-1}\frac{1-A(g,g_0)/A(g_0,g_0)}{g-g_0} \Big) .
\end{equation}
Hence we have 
\begin{equation}\label{r_near0:scaledwidth_remainder}
\begin{aligned}
\frac{g_0F_1(g)+gF_2(g,g_0)}{g(g-g_0)} = & 
\frac{1}{\sqrt{g_1-g}\sqrt{B(g,g_0)}}\bigg( 
\frac{g^\frac{k-3}{2}H(g,g_0)}{\sqrt{1-g^{k-1}} + \sqrt{A(g,g_0)}}
\\&\qquad
+ \frac{C(g,g_0)}{g_0\big(g_0{}^\frac{k-3}{2}g\sqrt{A(g,g_0)} + \sqrt{A(g_0,g_0)} g^\frac{k-1}{2}\big)g} 
\bigg) . 
\end{aligned}
\end{equation}
We will now use this expression to estimate the size of the corresponding integral term on the right-hand side of equation \eqref{r_near0:scaledwidth_integral},
which is given by the sum of the integrals
\begin{align}
& \int_{g_0}^{g_1} 
\frac{g^\frac{k-3}{2}H(g,g_0)}{\sqrt{g_1-g}\sqrt{B(g,g_0)}\big(\sqrt{1-g^{k-1}} + \sqrt{A(g,g_0)}\big)}\, dg ,
\label{r_near0:scaledwidth_remainder_integral:H}
\\
& \int_{g_0}^{g_1} 
\frac{C(g,g_0)}{g_0\sqrt{g_1-g}\sqrt{B(g,g_0)}\big(g_0{}^\frac{k-3}{2}g\sqrt{A(g,g_0)} + \sqrt{A(g_0,g_0)} g^\frac{k-1}{2}\big)g} .
\label{r_near0:scaledwidth_remainder_integral:C}
\end{align}
It is hard to obtain a good estimate that holds for arbitrary $k>1$,
so we will examine two special cases $k=2,3$
for which both of the integrals can be evaluated explicitly in terms of elementary functions, 
with $g_0/g^* =\epsilon^\frac{2}{k-1} \ll 1$
and $g_1\simeq 1-\tfrac{2}{k-1}g^*\epsilon^\frac{2}{k-1}$. 
(The case $k=\tfrac{3}{2}$ can be be evaluated in terms of elliptic functions.)

For $k=2$ and $k=3$, we get
\begin{equation}
\begin{aligned}
\int_{g_0}^{g_1} \frac{(g_0F_1(g)+gF_2(g,g_0))\,dg}{g(g-g_0)} 
& \simeq \sqrt{g_0} \big( \ln(g_0) +2\ln 2 \big)  +\arccos(1-4g_0)
\\
& \simeq \sqrt{g_0}\ln(g_0) + O(\sqrt{g_0}) ,
\end{aligned}
\end{equation}
and 
\begin{equation}
\begin{aligned}
\int_{g_0}^{g_1} \frac{(g_0F_1(g)+gF_2(g,g_0))\,dg}{g(g-g_0)} 
& \simeq g_0 \big( \ln(1-2g_0) +\ln 2 \big) +\arccos(1-g_0) 
\\
& \simeq \sqrt{2g_0} + O(g_0) . 
\end{aligned}
\end{equation}
In these two cases, we see that the sum of the two remainder integrals \eqref{r_near0:scaledwidth_remainder_integral:H} and \eqref{r_near0:scaledwidth_remainder_integral:C} 
vanishes as $g_0\to 0$.

Finally, 
combining the asymptotic expansion of the integral part of the width integral \eqref{width:g}
and the algebraic part of the width integral \eqref{width:g},
we have 
\begin{equation}
\hat\ell \simeq \sqrt{\tfrac{k(k+1)}{2}} \tfrac{\pi}{k-1} + o(\epsilon) . 
\end{equation}

\section{Derivation of KdV equation in the weakly nonlinear regime}\label{appendix:KdV}

Starting from the LWHC wave equation \eqref{LWCE-u}
and using the relation between the amplitude $u(t,x)$ and the continuum limit $U(t,x)$ of the particle displacement \eqref{U},
we introduce a scaled amplitude by putting 
\begin{equation}
U(t,x) = \epsilon v_0 Z(\tau,\xi) 
\end{equation}
with $\epsilon\ll 1$ being an expansion parameter, 
where $v_0$ is the background strain \eqref{initialstrain},
and where 
\begin{equation}\label{2scale}
\tau= \epsilon_2 t,
\quad
\xi = \epsilon_1 (x-c_0t)
\end{equation}
are a scaled time variable and a scaled space variable 
with respect to a reference frame moving with the sound speed \eqref{v0c0rel}. 
Here $\epsilon_1$ and $\epsilon_2$ are parameters that will be subsequently related to $\epsilon$. 
Note that the amplitude $u(t,x)$ is given by 
\begin{equation}\label{w-kdv}
u(t,x)/v_0 = \epsilon Z(\tau,\xi) -x . 
\end{equation}

We will now show that an expansion of the LWHC wave equation \eqref{LWCE-u}
leads to the KdV equation in potential form for $Z(\tau,\xi)$ 
with an appropriate choice of $\epsilon_1$ and $\epsilon_2$ in powers of $\epsilon$. 

To proceed, we substitute expression \eqref{w-kdv} into the LWHC wave equation \eqref{LWCE-u}
and use the relations
\begin{equation}
\begin{aligned}
& u_x/v_0 = \epsilon \epsilon_1 Z_\xi -1 ,
\quad
u_{xx}/v_0 = \epsilon \epsilon_1{}^2 Z_{\xi\xi}, 
\quad
u_{xxx}/v_0 = \epsilon \epsilon_1{}^3 Z_{\xi\xi\xi}, 
\quad
u_{xxxx}/v_0 = \epsilon \epsilon_1{}^3 Z_{\xi\xi\xi\xi}, 
\\
& u_t/v_0 = \epsilon (\epsilon_2 Z_\tau -\epsilon_1 c_0 Z_\xi),
\quad
u_{tt}/v_0 = \epsilon (\epsilon_2{}^2 Z_{\tau\tau} -2\epsilon_1\epsilon_2 c_0 Z_{\tau\xi} + \epsilon_1{}^2 c_0{}^2 Z_{\xi\xi}) . 
\end{aligned}
\end{equation}
Next we expand the resulting terms in a series in $\epsilon$,
yielding 
\begin{equation}\label{LHWC-expand}
\begin{aligned}
0=
& \epsilon \epsilon_1\epsilon_2 (2v_0/c^2) Z_{\tau\xi} 
+ \epsilon^2 \epsilon_1^3 (1-k)v_0^k Z_\xi Z_{\xi\xi}  
+ \epsilon \epsilon_1^4 \gamma v_0^k Z_{\xi\xi\xi\xi}  
\\& -\epsilon \epsilon_2{}^2 (v_0/c^2) Z_{\tau\tau} 
- \epsilon^2 \epsilon_1^5 \beta v_0^k Z_{\xi\xi}Z_{\xi\xi\xi}
+ \epsilon^3 \epsilon_1^6 \alpha v_0^k Z_{\xi\xi}^3
\\& +\text{ higher order terms }
\end{aligned}
\end{equation}
after we use expression \eqref{v0c0rel} for the sound speed $c_0$. 
The three terms in the first line of this equation \eqref{LHWC-expand}
correspond to the terms in the KdV equation in potential form for $Z$. 
Hence, we balance these terms by putting 
$\epsilon \epsilon_1\epsilon_2= \epsilon^2 \epsilon_1^3 =\epsilon \epsilon_1^4$,
which determines
\begin{equation}
\epsilon_1=\epsilon,
\quad
\epsilon_2=\epsilon^3 .
\end{equation}
Then we see that equation \eqref{LHWC-expand} becomes
\begin{equation}
0=
(2v_0/c^2) Z_{\tau\xi} + (1-k)v_0^k Z_\xi Z_{\xi\xi}  + \gamma v_0^k Z_{\xi\xi\xi\xi}  
+ O(\epsilon^2) . 
\end{equation}
Hence, in the limit $\epsilon\to 0$, we obtain 
\begin{equation}
(2v_0/c^2) Z_{\tau\xi} + (1-k)v_0^k Z_\xi Z_{\xi\xi}  + \gamma v_0^k Z_{\xi\xi\xi\xi}  \simeq 0
\end{equation}
which is the KdV equation for $Z_\xi$. 

The scaled amplitude $Z(\tau,\xi)$ can be directly expressed in terms of the strain variable \eqref{strain} by $v/v_0 = 1 - \epsilon^2 Z_\xi$,
and hence 
\begin{equation}
\epsilon^2 Z_\xi = 1-v/v_0 =-w
\end{equation}
where $w$ is the variable introduced in the expansion \eqref{smallv} 
for solitary waves in the weakly nonlinear regime
with $0<w\ll 1$.

\end{document}